\documentclass[]{aastex631}

\usepackage{amsmath}

\usepackage[normalem]{ulem}
\usepackage{enumitem}

\usepackage{blkarray}
\usepackage{mathtools}
\usepackage{cancel}

\usepackage{newunicodechar,graphicx} 
\DeclareRobustCommand{\okina}{%
  \raisebox{\dimexpr\fontcharht\font`A-\height}{%
    \scalebox{0.8}{`}%
  }%
}
\newunicodechar{ʻ}{\okina}

\defcitealias{Raboonik2024a}{P1}
\defcitealias{Raboonik2024b}{P2}

\usepackage{multirow}

\let\orgautoref\autoref

\renewcommand{\autoref}[1]{\def\equationautorefname{Equation}\orgautoref{#1}}

\newcommand*{\myeqref}[2][Eq.~]{%
  \hyperref[{#2}]{#1(\ref*{#2})}%
}
\def\equationautorefname#1#2\null{%
  Eq.#1(#2\null)%
}

\renewcommand{\leq}{\leqslant}  
  
\newcommand{\pow}[1]{^{#1}}
\newcommand{\poww}[1]{\pow{2}}
\newcommand{\powww}[1]{^{3}}

\newcommand{\bs}[1]{\boldsymbol{#1}}
\newcommand{\mathbfit}[1]{\bs{\mathit{#1}}}
\renewcommand{\a}{{\mathbfit{a}}}
\renewcommand{\b}{{\mathbfit{b}}}
\newcommand{\e}{\hat{\bs{e}}}
\newcommand{\g}{\mathbfit{g}}
\renewcommand{\k}{\mathbfit{k}}

\newcommand{\m}{\mathbfit{m}}

\renewcommand{\v}{{\mathbfit{v}}}
\newcommand{\x}{\mathbfit{x}}
\newcommand{\B}{{\mathbfit{B}}}

\newcommand{\E}{{\mathbfit{E}}}
\newcommand{\F}{{{{}_2F_3}}}
\newcommand{\X}{\mathbfit{X}}
\renewcommand{\P}{{\mathbfit{P}}}
\newcommand{\U}{\mathbfit{U}}

\newcommand{\D}{\mbox{\boldmath$\mathsf{D}$}}
\newcommand{\R}{\mbox{\boldmath$\mathsf{R}$}}
\newcommand{\Q}{\mbox{\boldmath$\mathsf{Q}$}}
\newcommand{\J}{\mbox{\boldmath$\mathsf{J}$}}
\newcommand{\Y}{\mbox{\boldmath$\mathsf{Y}$}}

\newcommand{\vdot}{{\mathbf{\cdot}}}

\newcommand{\grad}{\mbox{\boldmath$\nabla$}}

\renewcommand\div{\grad\vdot}

\newcommand{\alf}{Alfv{\'e}n}

\renewcommand{\a}{{\mathbfit{a}}}
\renewcommand{\b}{{\mathbfit{b}}}
\renewcommand{\k}{\mathbfit{k}}
\renewcommand{\v}{{\mathbfit{v}}}
\renewcommand{\P}{{\mathbfit{P}}}

\newcommand{\modeconvert}[3]{#2$\overset{#1}{\rightarrow}$#3}

\newcommand{\lrp}[1]{\left( #1 \right)}

\usepackage{xcolor}
\definecolor{forestgreen(web)}{rgb}{0.13, 0.55, 0.13}

\begin{document}

\title{Exact Nonlinear Decomposition of Ideal-MHD Waves Using Eigenenergies III: Gravity, Generalized Inhomogeneous Quasi-linear PDEs, Mode Conversion, and Numerical Implementation \footnote{Released on March, 1st, 2021}}

\author[0000-0002-6408-1829]{Abbas Raboonik}
\affiliation{The University of Newcastle,
University Dr, Callaghan,
NSW 2308, Australia}
\email{raboonik@gmail.com}

\author[0000-0002-1089-9270]{David I. Pontin}
\affiliation{The University of Newcastle,
University Dr, Callaghan,
NSW 2308, Australia}

\author[0000-0002-8259-8303]{Lucas A. Tarr}
\affiliation{National Solar Observatory, 22 Ohi\okina{}a Ku St, Makawao, HI, 96768}



\begin{abstract}
Precise tracking and measurement of the energy carried by the individual magneto\-hydro\-dynamic (MHD) modes has important implications and utility in astrophysical and laboratory plasmas. Previously, this was only achievable in limited linear MHD cases in the $\beta \ll 1$ or $\beta \gg 1$ regimes. In a series of papers, of which this is the third, we introduced the Eigenenergy Decomposition Method (EEDM) and derived exact analytical expressions for the modal energy components--called eigenenergies--of nonlinear 3D disturbances governed by the homogeneous ideal-MHD equations. Here, we extend the method to inhomogeneous ideal-MHD by introducing a source term accounting for gravity, and provide detailed guidelines for applying the decomposition scheme to any general inhomogeneous quasi-linear PDEs that possess a globally conserved quantity, beyond the realm of MHD. Furthermore, we show that the eigenenergies can be used to locate and measure nonlinear mode conversions, which is an additional feature of the method. Finally, we provide well-categorized context for the application of the method to simulations and discuss the possible numerical inaccuracies that may inevitably arise due to discretization. This paper provides a more mature description of the method and its interpretation, and is recommended as the starting point for readers unfamiliar with the method.
\end{abstract}



\section{Background and preliminary equation}
A fundamental process in astrophysical and laboratory plasmas is the propagation of waves, which carry energy and information through the plasma. 
On large scales the plasma is often well described using the framework of magneto\-hydro\-dynamics (MHD), within which the fundamental wave modes are the \cite{Alfven1942} waves and the slow and fast magneto\-acoustic waves \citep{banos1955}. There is a long history of studying the behavior of these waves in the Sun's corona, driven by the facts that (i) wave damping may contribute to explaining the anomalously high temperature of the Sun's atmosphere \citep[e.g.][]{vandoorsselaere2020}, and (ii) coronal seismology techniques have been developed that allow physical variables to be inferred through observations of these waves \citep[e.g.][]{demoortel2012,nakariakov2020}.

The MHD wave modes have unique physical properties, which determine how they interact with the plasma as they propagate. To this end, breaking down a general composite MHD wave into its modal components is the key to unlocking the hidden information underlying the wave, offering deep insights into the nature of plasma evolution. Previous attempts have been made to accomplish this task \citep{RosBogCar02aa,Cally2015,Tarr2017,Raboonik2019,Raboonik2021,yadav2022}. However, all the proposed methods are inexact, limited to extreme plasma-$\beta$\footnote{The plasma-$\beta$ is a dimensionless quantity defined as the ratio of the thermal pressure to magnetic pressure.} regimes ($\beta \ll 1$ or $\beta \gg 1$), of which many are only valid in sub-3D linear MHD and/or reliant on unphysical proxies used as ``surrogates'' for the MHD modes.

This paper builds on our previous work, started in \citet{Raboonik2024a} (henceforth referred to as \citetalias{Raboonik2024a}) and \cite{Raboonik2024b} (henceforth referred to as \citetalias{Raboonik2024b}), in which we developed an exact mathematical framework for uniquely analyzing general 3D nonlinear MHD evolutions in terms of the characteristic wave modes. The principal aim of this study is to extend our previous work by including the gravity force in the equations.

One way to understand the generalization of the concept of linear MHD wave modes to a nonlinear evolution is to consider the MHD equations as an eigensystem \citep{Jeffrey1964}. This can be done in the ideal (perfectly conducting) limit, where the MHD equations in the absence of external effects (such as gravity) can be written in the homogeneous quasi-linear form as
\begin{equation}\label{eq:mhd}
    \dot{\bs{P}} + M_q \partial_q \bs{P} = 0,
\end{equation}
in which the overdot represents the Eulerian time derivative, and Einstein's summation rule is implied (and assumed throughout the paper unless stated otherwise). Here $\bs{P} = \lrp{\rho, v_x, v_y, v_z, B_x, B_y, B_z, p}^T$, known \emph{a priori}, is the $8\times 1$ state (or solution) vector, wherein $\rho$ is the mass density, $B_q$ and $v_q$ are respectively the $q$-components of the magnetic field and plasma velocity in Cartesian coordinates $q \in \lrp{x,y,z}$, $p$ is the thermal pressure, and $M_q$ are the globally non-defective $8\times8$ local flux matrices (given in Appendix \ref{sec:AppendixA}) describing all the internal forces. Therefore, upon diagonalizing $M_q$, we find that the system has eight eigenvalues in each direction $q$. These are the plasma velocity (repeated with the algebraic multiplicity of two), and the differences ($\pm$) between the plasma velocity and the Alfv\'en, slow, and fast speeds \citep{roe1996}. Each of these eigenvalues is a ``characteristic speed'' in the Eulerian picture, with the corresponding eigenvector describing the associated characteristic curve (in the state space defined by ${\bf P}$). Such a formulation of an MHD evolution in terms of the contributions from these different characteristics has in the past been exploited in developing numerical solvers \citep{HarLax83,roe1996} and implementing boundary conditions (BCs) \citep{Nakagawa1980, cimino2016,Tarr2024}. 

In \citetalias{Raboonik2024a} and \citetalias{Raboonik2024b} we introduced a method for understanding the nonlinear evolution of ideal-MHD disturbances in the absence of gravity, by examining the variation of energy density in space and time, and decomposing this energy variation into the contributions from each of the eight characteristic modes in each of the three spatial directions (the ``Eigenenergy Decomposition Method'', or EEDM). In \citetalias{Raboonik2024a}, the core mathematical method was described and the results of 1.5D and 3D simulations were presented to demonstrate that the method is capable of uniquely identifying propagating wave modes of known types in those nonlinear situations. Then in \citetalias{Raboonik2024b} we refined the method by removing local singularities that appeared where certain characteristic speeds became zero \citep[building on][]{roe1996}. We also addressed the nature of the zero-frequency modes--the two so-called ``pseudo-advective'' modes with a repeated eigenvalue equal to the plasma velocity--illustrating that they independently describe the advection of a component of the magnetic energy density, and a component of the internal energy density due to the plasma entropy.

One restriction of the EEDM as presented in \citetalias{Raboonik2024a} and \citetalias{Raboonik2024b} is that it applies only to a perfectly ideal MHD evolution without resistive or viscous dissipation, and with the energy equation being provided by the adiabatic assumption. This is because the MHD equations in this form are \emph{hyperbolic}, which is required in order to write them in the form of Equation~\ref{eq:mhd}. Another crucial piece of physics that is excluded from this equation is gravity. The goal of this paper is to extend the EEDM to include a general external gravitational force.

The method that we outline here opens up the potential for incorporating more physics to the ideal-MHD case in future works, by following the guidelines described herein. In the meantime, however, the current method can still provide estimates in the case of weakly non-ideal plasmas, or exact results in the ideal-MHD regions of plasmas wherein non-ideal effects are confined to spatio\-temporally localized regions.
We note, further, that the method introduced here may be broadly applicable in different fields (i.e., when using a characteristic description of different sets of equations). In such cases, as when gravity is introduced into the MHD equations (see the following section), it is common to deal with forcing terms that render the equations into an inhomogeneous quasi-linear system. Since \citetalias{Raboonik2024a}, our understanding of the EEDM, its strengths, and numerical implementations has significantly matured. For this reason, we recommend the reader approach the paper series backwards in time, i.e., start with this paper as a first introduction to the method, followed by \citetalias{Raboonik2024b} and \citetalias{Raboonik2024a} for complementary details.

In Section~\ref{sec:level2} we present the mathematics of the extended method. Then in Section~\ref{sec:level3} we illustrate its implementation in MHD simulations. Finally, we present our conclusions in Section~\ref{sec:conc}.

\section{\label{sec:level2}The Inhomogeneous eigenenergy Decomposition Method}
In this section, we present a step-by-step guide to deriving the mode-decomposed energy density components of an ideal plasma under the external force of gravity. Although here we will only discuss the ideal-MHD PDEs, these guidelines can be applied to any other quasi-linear PDEs (homogeneous or inhomogeneous) that describe a system constrained by a global conservation law in order to break down the conserved quantity into its modal components. As the term ``eigenenergy'' suggests, the globally conserved MHD variable to be decomposed in this paper is the total energy density. From here on out, following our convention from \citetalias{Raboonik2024a} we shall assume that all instances of ``energy'' refer to energy \emph{density}.

Assuming gravity acts only as an external force on an ideal plasma, i.e., ignoring self-gravity, the set of Eulerian gravitational MHD equations can be written in the \emph{generalized inhomogeneous quasi-linear}  form as follows
\begin{eqnarray}\label{eq:MHD}
    \dot{\bs{P}} + M_q \partial_q \bs{P} = \bs{S},
\end{eqnarray}
whose only difference from Equation~\ref{eq:mhd} is the addition of the source or forcing vector $\bs{S} = \lrp{0,-\bs{\nabla}\phi,0,0,0,0}$, with $\phi$ representing the gravitational potential. Generally, $\bs{S}$ may vary both spatially and temporally so long as it describes a pure external force, i.e., its evolution is independent of the state vector $\bs{P}$ (whereas the evolution of $\bs{P}$ depends upon $\bs{S}$). All the coupling mechanisms mediated by the internal forces are encoded in $M_q$.

The core idea behind the Eigenenergy Decomposition Method (EEDM) is to break down the time variations of the total energy, given by
\begin{equation}\label{eq:totalEn}
    \dot{E}_\text{tot} = \frac{1}{2} \dot{\rho} v^2  + \rho \bs{v}\vdot\dot{\bs{v}} + \frac{\dot{p}}{\gamma - 1} + \frac{1}{\mu_0} \bs{B}\vdot\dot{\bs{B}} + \dot{\rho}\phi,
\end{equation}
in terms of the individual energy contributions from all the modes supported by Equation~\ref{eq:MHD}. 
For simplicity, we take $\phi$ to be time-independent, but note that by following the guidelines below the method can be readily extended to incorporate the time-evolution of gravity by adding $\rho \dot{\phi}$ to the equation above. Combining Equations~\ref{eq:MHD} and \ref{eq:totalEn}, the decomposition can be accomplished in two separate steps. 

First, similar to the homogeneous case covered in \citetalias{Raboonik2024a} and \citetalias{Raboonik2024b}, we use the eigensystems of $M_q$ (whose eigenvectors provide unique signatures of the MHD eigenmodes) to derive the time evolution of the energy components carried by each eigenmode, called the eigenenergy variations.
Second, once the eigenenergy variations are known, the remaining energy variations due to any other modes induced by $\bs{S}$ can be derived by a simple subtraction. These steps are described in more detail below. More physics may be readily incorporated by modifying $\bs{S}$ to account for other purely external effects, or $M_q$ for other internal coupling mechanisms, provided that the ensuing PDEs remain quasi-linear and possess an eigensystem.

We begin by diagonalizing the flux matrices according to $M_q = R_q \Lambda_q L_q$, wherein $L_q$ and $R_q$ ( $= L_q^{-1}$), given in Appendix~\ref{sec:AppendixB} \citep[see][originally adapted from \cite{roe1996}]{Raboonik2024b}, are the fully analytical biorthonormal left and right eigenmatrices, respectively, and $\Lambda_q = \text{diag} (v_q, v_q, v_q-\|a_q\|, v_q+\|a_q\|, v_q-c_{\text{s},q}, v_q+c_{\text{s},q}, v_q - c_{\text{f},q}, v_q+c_{\text{f},q})$ is the diagonal matrix of eigenvalues. 
The elements of $\Lambda_q$, composed of the three MHD characteristic speeds and the bulk velocity, are the familiar $q$-directed \alf{} speed $a_q = B_q / \sqrt{\mu_0 \rho}$ with $\mu_0$ signifying the permeability of the vacuum, and the fast/slow speed defined by $c_{\text{f/s},q} = \lrp{a^2 + c^2 \pm \sqrt{a^4 + c^4 + 2 c^2(a^2 - 2a_q^2)}}^{1/2}/\sqrt{2}$, where $a = \|\bs{a}\|$, and $c =  \sqrt{\gamma p / \rho}$ is the adiabatic sound speed with $\gamma$ denoting the heat capacity ratio.
Consequently, we may rewrite Equation~\ref{eq:MHD} in the diagonalized form according to  
\begin{equation}\label{eq:Pdot}
    \dot{\bs{P}} = \bs{S} - R_q \bs{\mathcal{L}}_q,
\end{equation}
wherein $\bs{\mathcal{L}}_q = \Lambda_q L_q \partial_q \bs{P}$ is the \emph{mode-decomposed displacement eigenvector}.
Importantly, the significance of $\bs{\mathcal{L}}_q$ is that each of its eight elements $l_{m,q}$ identifies a unique eigenmode $m$ of the \emph{homogeneous} quasi-linear system in the direction $q$, encoding all the internal coupling effects.

The components of Equation~\ref{eq:Pdot} give the time variation of each of the state variables, written as a sum over contributions from the different modes of the system. Substituting these mode-decomposed elements of $\dot{\bs{P}}$ from Equation~\ref{eq:Pdot} into Equation~\ref{eq:totalEn}, we arrive at the eigenenergy decomposed version of $\dot{E}_\text{tot}$ comprised of a total of twenty-five (eight eigenmodes in each of the three directions $q\in\{ x,y,z\}$ plus an additional purely gravitational non-eigen mode) linearly independent terms. This can be written in terms of a linear functional summation as follows
\begin{equation}\label{eq:totalEnDecomp}
    \dot{E}_\text{tot} =  w_{m,q} l_{m,q} + \dot{E}_{\text{g}},
\end{equation}
where $m = \left[1..8\right]$ denotes the eigenmode number, $w_{m,q} = f\lrp{R_q, \phi}$ is the weight function describing the couplings between all the internal and external forces, the product $w_{m,q} l_{m,q} = \dot{E}_{m,q}$ (no summation) defines the individual eigenenergy variations associated with the eigenmodes of the inhomogeneous system, and the subscript ``g'' represents an additional independent \emph{driven} mode due exclusively to the external force of gravity. In other words, and as we shall see, $\dot{E}_g$ is indeed entirely uncoupled from the internal forces, purely capturing the work done by (or against) gravity alone. Thus, to summarize, $l_{m,q}$ which serve as eigenmode identifiers, are the components of the displacement eigenvectors $\bs{\mathcal{L}}_q$ corresponding to the homogeneous case (same as those of \cite{Raboonik2024b}), the weight coefficients $w_{m,q}$ encode the combined effects of the internal and external forces, and $\dot{E}_g$ accounts for the (non-eigen) mode driven solely by the source vector $\bs{S}$.

To further expound on the method, note that purely external forces do not change the characteristics of the quasi-linear system, whereas any modification to $M_q$ induces characteristic alterations. Therefore, the exact expressions for $\dot{E}_{m,q}$ can be derived by tracing the unique identifiers $l_{m,q}$ in the post-decomposition expansion of  $\dot{E}_{\text{tot}}$, and recording their coefficients as $w_{m,q}$. The new driven mode corresponding to $\dot{E}_{\text{g}}$ is then sifted out by computing the leftover terms that balance out Equation~\ref{eq:totalEnDecomp}, i.e., $\dot{E}_{\text{g}} = \dot{E}_{\text{tot}} -  w_{m,q} l_{m,q}$, where $\dot{E}_{\text{tot}}$ is given by Equation~\ref{eq:totalEn}. Moreover, $\dot{E}_{\text{g}}$ itself can be readily decomposed into the three spatial components, i.e., $\dot{E}_{\text{g}} = \sum_q \dot{E}_{\text{g},q}$. Thus, for this specific form of $\bs{S}$, there are a total of nine modes (eight eigenmodes $m = 1, 2, ..., 8$, plus the additional driven mode $m = 9$) in each spatial direction.

Due to the unique physical nature of these modes, we classify them in what follows as the field-divergence (branch name shorthand div; $m = 1$) and entropy (ent; $m = 2$) \emph{pseudo-advective (PA)} eigenmodes \cite[see][for a full discussion of these modes]{Raboonik2024b}, the reverse ($-$) and forward ($+$) \alf{} (A; $m = 3, 4$), slow (s; $m = 5, 6$), and fast (f; $m = 7, 8$) \emph{natural} eigenmodes, and the gravity (g; $m = 9$) \emph{driven PA} mode.
The different terminologies here reflect the facts that (\emph{i}) the three PA modes are degenerate and non-propagating (zero-frequency), rendering them strictly advective in nature, meaning that they describe the transport of a portion of the total energy along the flow at the rate of $v_q$, (\emph{ii}) $m \in [1..8]$ are all associated with the eigensystems of $M_q$, and hence are the eigenmodes, while (\emph{iii}) the driven PA mode $m = 9$ is induced by $\bs{S}$, and hence is not an eigenmode. Note that we have termed the collection of the A, s, and f branches ($m \in [3..8]$) the natural eigenmodes.

Table~\ref{tab:modes} presents a summary of all the modes mentioned above, their name classification, and important attributes. 
Having introduced the formal nomenclature and distinguished between the eigenmodes and the driven mode, we are now in a position to use ``mode'' and ``eigenmode'' interchangeably to refer to any of the nine $m$'s listed in this table as a matter of convenience (even though $m = 9$ is technically not an eigenmode). Accordingly, we will call $E_{9,q}$ an eigenenergy despite its (non-eigen) driven origin.

\subsection{Inhomogeneous EEDM Equations}
Following the derivation procedure explained above, the component eigenenergy variations associated with fully nonlinear 3D gravitational ideal-MHD disturbances satisfying Equation~\ref{eq:totalEnDecomp}, henceforth referred to as the inhomogeneous EEDM equations, are found to be (no summation over the repeated index $q$)
\begin{subequations}\label{eq:EEDMNew}
    \begin{equation}\label{eq:div}
        \dot{E}_{\text{div},q} = -\frac{B_q \partial_q{B_q}}{\mu_0} v_q,
    \end{equation}
    
    \begin{equation}\label{eq:ent}
        \dot{E}_{\text{ent},q} = -\frac{\lrp{v^2 + 2 \phi} \lrp{c^2 \partial_q{\rho} - \partial_q{p}}}{2 c^2} v_q,
    \end{equation}
    
    \begin{equation}\label{eq:alfNew}
        \dot{E}_{\text{A},q}^{\mp} = - \big(\sqrt{\mu_0 \rho}\bs{v}\bs{\times}\bs{\mathcal{B}}_q\big)_q  \Big(\big(\sqrt{\mu_0 \rho}\partial_q{\bs{v}} \pm s_{q}\partial_q{\bs{B} }\big)\bs{\times}\bs{\mathcal{B}}_q\Big)_q \dfrac{\lrp{v_q \mp \|a_q\|}}{2 \mu_0},
    \end{equation}
    
    \begin{eqnarray}\label{eq:sloFasNew}\begin{array}{c}
        \dot{E}_{\text{s/f},q}^{\mp} = -\lrp{\mathcal{S}_\text{s/f} \alpha_{\text{f/s},q} \Big(\pm s_q c_{\text{f/s},q} \sqrt{{\mu }_{0}\rho } \bs{v} + c \bs{B}\Big)\vdot\bs{\mathcal{B}}_q + \sqrt{{\mu }_{0}\rho }\alpha_{\text{s/f},q}\Big(\pm c_{\text{s/f},q} v_q - \frac{c^2}{\gamma - 1} - \frac{1}{2} v^2 - \phi\Big) }\\ \hspace{1.49cm}\times \,\lrp{\mathcal{S}_\text{s/f}\alpha_{\text{f/s},q}  \Big(\pm s_q c_{\text{f/s},q} \sqrt{\mu_0 \rho} \partial_q \bs{v} + c \partial_q \bs{B}\Big)\vdot \bs{\mathcal{B}}_q +\sqrt{\mu_0 \rho}\alpha_{\text{s/f},q}\Big(\pm c_{\text{s/f},q}\partial_q v_q - \dfrac{1}{\rho} \partial_q p\Big)}\dfrac{ \lrp{v_q \mp c_{\text{s/f},q}}}{2 \mu_0{c}^{2}},\end{array} 
    \end{eqnarray}

    \begin{equation}\label{eq:dotEg}
        \dot{E}_{\text{g},q} = -\rho (\partial_q \phi) v_q.
    \end{equation}
\end{subequations}
Here $\alpha_{\text{s/f},q} = \lrp{\mathcal{S}_\text{s/f} \big(c_{\text{f/s},q}^2 - c^2\big)/\big(c_{\text{f},q}^2 - c_{\text{s},q}^2\big)}^{1/2}$
with $\mathcal{S}_\text{s} = 1$ and $\mathcal{S}_\text{f} = -1$ \citep[thereby satisfying the identity $\alpha_{\text{s},q}^2 + \alpha_{\text{f},q}^2 = 1$; see][]{roe1996}, $s_q = \text{sgn}(a_q)$, and $\bs{\mathcal{B}}_q$ is the $q$-th row of the $3\times3$ matrix $\mathcal{B}_{q,q'} = \beta_{q'\perp q}\lrp{1 - \delta_{q,q'}}$ with
\begin{equation}\label{eq:beta}
        \beta_{q' \perp q} = \begin{cases} 
      a_{q'}/a_{\perp q} & a_{\perp q} \neq 0 \\
      \frac{1}{\sqrt{2}} & \text{otherwise} 
   \end{cases},
\end{equation}
wherein $q' \perp q$ denotes parallel to $q'$ and perpendicular to $q$ (e.g., $\beta_{x \perp y} = a_x/\sqrt{a_x^2 + a_z^2}$; not to be confused with the plasma-$\beta$). 
Note the purely magnetic nature of the field divergence and \alf{} components, the purely acoustic (non-magnetic) nature of the entropy and gravity components, and the mixed magnetic-acoustic (magneto\-acoustic) nature of the slow and fast components, as expected. Note further the elegant separation of the purely magnetic (the terms involving $\alpha_{\text{f/s},q}$) and the purely acoustic (the terms involving $\alpha_{\text{s/f},q}$) parts of the slow and fast components (more on this as well as the mathematical nature of the $\alpha$ parameters in Section~\ref{sec:simI}).

\begin{table}[]
\begin{tabular}{c|c|c|c|c|c}
$m$ & Name & Branch shorthand & Characteristic direction $(d)$ & Nature/Collective name & Origin \\ \hline \hline
1 & Field-divergence & div & \multirow{2}{*}{Unidirectional} & \multirow{2}{*}{Non-propagating/PA} & \multirow{8}{*}{Eigenmode} \\ \cline{1-3} 
2 & Entropy & ent & & & \\ \cline{1-5}
3 & Reverse \alf & \multirow{2}{*}{A} & $-$ & & \\ \cline{1-2} \cline{4-4}
4 & Forward \alf & & $+$ & Propagating & \\ \cline{1-4}
5 & Reverse slow & \multirow{2}{*}{s} & $-$ & and & \\ \cline{1-2} \cline{4-4}
6 & Forward slow & & $+$ & non-propagating/Natural & \\ \cline{1-4}
7 & Reverse fast & \multirow{2}{*}{f} & $-$ & & \\ \cline{1-2} \cline{4-4}
8 & Forward fast & & $+$ & & \\ \hline
9 & Gravity & g & N/A (Non-characteristic) & Non-propagating/PA & Driven mode
\end{tabular}
\caption{The classification and nomenclature used in this paper to identify all the gravitational ideal-MHD modes in each spatial direction $q$. The MHD branches listed in the third column refer to the eigenmode names stripped of their characteristic direction (where applicable). The shorthand ``PA'' stands for pseudo-advective.}\label{tab:modes}
\end{table}

The decomposed total energy itself can be derived from
\begin{equation}\label{eq:totalEnergyDecomposition}
    E_\text{tot} - E_0 = \sum_{q\in\lrp{x,y,z}}\sum_{m = 1}^9 E_{m,q},
\end{equation}
in which $E_0 = E_\text{tot}(\bs{x},t_0)$ is the initial total energy and
\begin{equation}\label{eq:eigenenergy}
    E_{m,q} =\int_{t_0}^{t}{\dot{E}_{m,q}\,dt'}, \quad \text{with }E_{m,q}\bigg|_{t = t_0} = 0
\end{equation}
are the eigenenergies. Thus, the instantaneous decomposition of the energy depends upon the initial energy $E_0$, or more precisely, upon the net energy difference $\Delta{E_\text{tot}} = E_\text{tot} - E_0$. As a result, for two different initial timestamps $t_0 \neq \tau_0$, we have $E_{m,q}(t;t_0) \neq E_{m,q}(t;\tau_0)$. 
The initial value of zero enforced on all the eigenenergies $E_{m,q}$ at $t_0$ in the time integral above is to ensure the logical requirement $\Delta E_\text{tot}(t = t_0) = 0$ regardless of $E_0$.
Note that the eigenenergies defined in Equations~\ref{eq:EEDMNew} are functions of both space and time.

In Section~\ref{sec:level3} we illustrate how the EEDM as described above can be used to decompose the energy in a nonlinear MHD evolution. But first, in the remainder of this section let us discuss some further aspects of the method itself, starting with mode conversion.

\subsection{EEDM and Nonlinear Mode Conversion}\label{sec:modeConversion}
Equation~\ref{eq:eigenenergy} assumes no precondition on the initial state of the plasma and remains valid for any $E_0$, even if it describes an already dynamic initial state. In other words, in defining the eigenenergies, $E_0$ can be composed of some mixture of initial eigenenergies of its own, of which we assume no knowledge\footnote{An exact decomposition of a dynamic initial state $E_0$ based on its eigenenergy variations would itself require knowledge about the system prior to $t_0$. Thus, any EEDM-like nonlinear decomposition scheme would have to be independent of the decomposition of $E_0$ to avoid an infinite regression and/or limiting it to only fully static initial states.}. This independence of the distribution of the initial eigenenergies (that make up $E_0$) is built into the EEDM through the boundary values of the time-integral in Equation~\ref{eq:eigenenergy}, i.e., setting $E_{m,q}(t_0) = 0$. As a defining consequence, the eigenenergies $E_{m,q}$ (which sum up to $\Delta{E_\text{tot}}$) can take negative values.

At first glance, this may seem counter-intuitive. However, note that the nonlinear evolution of the plasma can lead to a local decrease of the total energy compared to the initial state, i.e., $\Delta{E_\text{tot}} < 0$. 
Therefore, the eigenenergies, which describe the evolution of $E_\text{tot}$, must be able to both add and subtract energy to and from the system as it evolves. A familiar example is the relaxation of a magnetic twist, where the initial total energy concentrated in the twist would propagate away from this location upon relaxation, leaving behind a region of negative $\Delta{E_\text{tot}}$, which would inevitably be described by some negative eigenenergies. More precisely, and guided by the results of this paper and \citetalias{Raboonik2024b}, we may interpret any negatively valued eigenenergies as an indication of mode conversion. This can be explained through two thought experiments as follows.

\subsubsection{Type 1 conversion: propagating-to-propagating}
Consider a simple 1D propagating wave along the $z$-axis, where the initial energy decomposition is somehow known \textit{a priori} to be precisely $E_{m,q}(t_0) = 0$ for all $m$ except for, say, $m = 8$. This means that the entire initial total energy is stored in the forward fast branch propagating at $c_{\text{f},z}$, i.e., $E_{8,z}(t_0) = E_0$. In light of this information, we may modify the lower boundary values of the time-integral of Equation~\ref{eq:eigenenergy} to reflect this fact, which in turn imposes a modification on Equation~\ref{eq:totalEnergyDecomposition} according to $E_\text{tot} = \sum_{m} E_{m,z} > 0$ (since the initial energy is now integrated into the eigenenergies). Now, suppose that due to some coupling effects, this fast energy is entirely transformed into slow energy at some time $t$ and location $\bs{x}$ and propagated away at the new rate $c_{\text{s},z}$ (as is the case in fast-slow mode conversion at the equipartition level $z_\text{eq}$, where $a = c$). In that case, as part of the post-conversion partitioning of the energy we would have $E_{6,z}(t) = E_0$ and $E_{m,z}(t) = 0$ for all other modes $m \neq 6$.

Now, how would the eigenenergies have changed in this scenario had we not known the initial energy decomposition \emph{a priori}, i.e., had we followed Equation~\ref{eq:eigenenergy} choosing $E_{m,z}(t = t_0) = 0$ for all $m$? The answer is that the fast-slow transformation would have resulted in a negative value (instead of zero) appearing in the fast branch (propagating at the fast speed), i.e., $E_{8,z}(t) = -E_0$, and a positive value in the slow branch (propagating at the slow speed), i.e., $E_{6,z}(t) = 2E_0$, so that $E_{6,z}(t) + E_{8,z}(t) = E_0$ due to conservation of energy. We shall term this kind of conversion from one propagating mode to another the Type 1 conversion, and indicate it by the overscripted arrow $\overset{1}{\rightarrow}$.

\subsubsection{Type 2 conversion: propagating-to-non-propagating}
As another thought experiment, suppose a propagating nonlinear fast pulse is traveling through a segment of a structured plasma that is initially at rest. As the fast wavefront travels through unperturbed regions governed by some local balance between the internal, magnetic, and gravitational energies, it disrupts the initial local equilibrium $E_0$, causing a redistribution of the energies as well as injecting some kinetic energy, thereby enforcing a reconfiguration of the background state. Since the eigenenergies capture different parts of the four forms of energy (kinetic, gravitational, magnetic, internal), any redistribution of energy will be described by a change in the eigenenergies
(which all assume an initial value of zero according to Equation~\ref{eq:eigenenergy}).

Let us focus on a reconfigured region of volume $V$ in the wake of the fast pulse that after some time $\delta t$ has settled into another equilibrium state $E_1$.
Thus, for all $x, y, z \in V$ we have $\Delta E_\text{tot} = E_1 - E_0 = F(x,y,z)$, where $\dot{F} = 0$ due to the new equilibrium. Since $F$ is a time-constant describing a \emph{non-propagating} state, we may deduce the following about the nature of its corresponding eigenenergies. First, all the eigenenergies associated with $F$ must be also non-propagating, i.e., of the PA nature, which according to \citetalias{Raboonik2024b} can include the natural modes as well (also see Table~\ref{tab:modes}). Second, some of the eigenenergies may be negatively valued such that the summation correctly recovers $F$. In other words, any new equilibrium state that has evolved from some initial state $E_0$ may be described by a combination of PA eigenenergies that admit negative values. The specific distribution of the eigenenergies depends upon how the initial distribution of the internal, magnetic, and gravitational energies was modified by the passage of the pulse. Put in the terminology of conversion, the fast wave, while predominantly transmitting through the plasma, would locally deposit some energy by mode-converting into some (positive and negative) PA eigenenergies whose net equals $F$. Such conversions would manifest themselves as (unchanging) eigenenergies trailing a pulse in its wake, and thus appear to co-move with it. This was previously observed in \citetalias{Raboonik2024a} and \citetalias{Raboonik2024b} and referred to as \emph{mode-degeneracy}. Let us term this kind of conversion from a propagating mode into non-propagating ones the Type 2 conversion, indicated by the overscripted arrow $\overset{2}{\rightarrow}$.

Overall, although the range of possible scenarios is not limited to these thought experiments and, as will be seen, the partitioning of the negative and positive eigenenergies may be much more complex in practice, nevertheless, as a principle of the EEDM, negative eigenenergies in Equation~\ref{eq:eigenenergy} are in general indicative of the two types of mode conversion described above.
To further emphasize, this principle is a direct product of initializing all the eigenenergies equally at zero by design, irrespective of $E_0$. Therefore, any negative values emerging in the eigenenergies can only originate from a redistribution of energy between the kinetic, internal, magnetic, and gravitational parts, which is expressible in terms of a redistribution of the eigenenergies, otherwise known as, mode conversion. Note that the existence of only two possible types of conversion is a direct consequence of the inherently binary classification of the eigenmodes into the propagating and non-propagating categories.

\subsection{Alternative Derivation of the Inhomogeneous EEMD Equations}
As an alternative way of arriving at Equations~\ref{eq:EEDMNew}, one may start from the already known homogeneous version of the EEDM equations (given in \citetalias{Raboonik2024b}), which we denote by $\dot{E}_{m,q}^\text{H} \forall m\in[1..8]$, and derive the inhomogeneous version, i.e., Equations~\ref{eq:EEDMNew}, through a syntactic variable replacement\footnote{Note that this is different from a change of variable.} as follows. To do so, we start by grouping together the first and the last terms of Equation~\ref{eq:totalEn} according to $\dot{E}_\text{tot} = \frac{1}{2} \dot{\rho} \lrp{v^2 + 2\phi}  + \rho \bs{v}\vdot\dot{\bs{v}} + \frac{\dot{p}}{\gamma - 1} + \frac{1}{\mu_0} \bs{B}\vdot\dot{\bs{B}}$. Note that syntactically, this has the same mathematical form as the homogeneous case, if we treat the terms in parentheses as an \emph{independent} variable $V^2 = v^2 + 2\phi$. Conversely, replacing $v^2$ with $V^2$ in the homogeneous total energy variation equation (see the Equation 7 of \citetalias{Raboonik2024a}) leads to the inhomogeneous version, i.e., Equation~\ref{eq:totalEn}. Thus, following this variable replacement rule, we may derive 
Equations~\ref{eq:EEDMNew} except for Equation~\ref{eq:dotEg} from $\dot{E}_{m,q}^\text{H}$ by replacing all the instances of $v^2$ in $\dot{E}_{m,q}^\text{H}$ with $V^2$ (i.e., with $v^2 + 2\phi$) according to
\begin{equation}
    \dot{E}_{m,q} = \dot{E}_{m,q}^\text{H}\bigg|_{v^2 \rightarrow v^2 + 2\phi}\quad\forall m\in[1..8].
\end{equation}
As before, Equation~\ref{eq:totalEnDecomp} can then be used to recover Equation~\ref{eq:dotEg}. 

While this approach may not apply in general, it offers the possibility of adding more physics to the base (homogeneous) case in a straightforward manner. In any case, the first detailed procedure presented here applies to any quasi-linear PDE of the form given by Equation~\ref{eq:MHD}.

\subsection{The Gravity Driven PA Mode}
To interpret the driven mode given by Equation~\ref{eq:dotEg}, note that gravity changes the total energy through doing (\emph{i}) direct work on the fluid as an independent external force, and (\emph{ii}) joint work in tandem with the internal forces.
Mathematically, this can be verified by decomposing the rate of change of the net gravitational energy denoted by $\dot{E}_\text{G}$ (note the difference between the subscripts G and $g$) as formulated by its associated flux vector\footnote{Alternatively, $\dot{E}_\text{G}$ can also be directly computed from Equations~\ref{eq:EEDMNew} and their homogeneous counterpart by isolating the gravitational terms through $\dot{E}_\text{G} = \Delta{E}_\text{tot} - \Delta{E}^\text{H}_\text{tot} = \sum_{m = 1}^{8}\sum_{q\in\lrp{x,y,z}}\lrp{\dot{E}_{m,q} - \dot{E}_{m,q}^\text{H}} + \dot{E}_g$.} $\bs{\mathcal{F}}_\text{G} = \rho \v \phi$ according to
\begin{align}\label{eq:totalGravEn}
    \dot{E}_\text{G} = -\div{\bs{\mathcal{F}}_\text{G}} &= -\div{\lrp{\rho \v}} \phi - \rho \v\vdot\bs{\nabla}\phi \nonumber \\
    &= \dot{\rho} \phi + \dot{E}_g.
\end{align}

Now, tracing $\dot{E}_g$ back reveals that it originates from the second term in Equation~\ref{eq:totalEn} (corresponding to the kinetic part of the total energy), and hence it describes (\emph{i}), i.e., how gravity directly affects the kinetic energy budget of the system. Specifically, $\dot{E}_g > 0$ if $\v\vdot\bs{\nabla}\phi < 0$ (i.e., flow towards the source of gravity) leading to an increase in the kinetic energy, and $\dot{E}_g < 0$ if $\v\vdot\bs{\nabla}\phi > 0$ (i.e., flow away from the source of gravity) resulting in a decrease in the kinetic energy. By contrast, the first term in the second line of Equation~\ref{eq:totalGravEn} describes (\emph{ii}), i.e., the other part of the gravitational energy that results from the coupling of gravity and the internal forces, and hence why it ultimately shows up in the $w_{m,q}$ coefficients.

\section{\label{sec:level3}Application to Simulations}
We shall here demonstrate the method by applying Equations~\ref{eq:EEDMNew} to two numerical simulations--named Simulations I and II following \citetalias{Raboonik2024a}\footnote{Unlike \citetalias{Raboonik2024a}, we no longer need to further break down Simulation I into two categories of high and low plasma-$\beta$, due to the gravitational stratification enforcing a variable $\beta$ profile even with a uniform magnetic field.}--mimicking the propagation of vortical nonlinear disturbances in gravitationally stratified model atmospheres driven at the solar photosphere. Note that this section serves primarily as an illustration of the numerical implementation of the EEDM, rather than a study of the solar atmosphere. The next few sections describe the simulations' initial conditions and results, while the details of the drivers are described in \autoref{sec:AppendixC}.

\subsection{Simulation Code and Numerical Scheme}
In common with \citetalias{Raboonik2024a}, the numerical simulations here are performed using Lare3d version 3.4.1 \citep{ARBER2001151}, which is a ``Lagrangian remap code'' that solves the nondimensional MHD equations using a finite difference scheme on a staggered grid, offering second-order accuracy in space and time.
In keeping with this choice, we shall nondimensionalize the EEDM equations by setting $\mu_0 = 1$ and treating all the other variables in Equations~\ref{eq:EEDMNew} as dimensionless. Thus, from here on out, any numerical values provided without specifying the physical units are to be taken as dimensionless. Furthermore, in all our numerical computations, the two staggered output variables of Lare3d, namely $\bs{v}$ and $\bs{B}$ \citep[see][]{ARBER2001151}, are first remapped onto a uniform cell-centered grid using 8- and 4-point averaging, respectively. All spatial partial differentiations are then done using the version 1.10.1 of SciPy's native cubic basis spline (B-spline) python routines ``splrep'' and ``splrev'', and all numerical time integrations are performed using the same library's ``cumtrapz'' routine \citep{2020SciPy-NMeth}.

Following in the footsteps of \citetalias{Raboonik2024a} (albeit using a slightly different notation for better clarity), we will quantify the numerical fidelity of the EEDM by how accurately Equation~\ref{eq:totalEnergyDecomposition} holds. In other words, the performance of a specific numerical implementation of the EEDM is determined by how well the \emph{pre-decomposition} net energy difference, i.e., $\Delta{E}_\text{tot} = E_\text{tot} - E_0$ ($ = E_\text{orig}$ in \citetalias{Raboonik2024a}), equals its \emph{post-decomposition} counterpart, which we denote by $\Delta{\mathcal{E}_\text{tot}} = \sum_{m,q}E_{m,q}$ ($ = E_\text{rec}$ in \citetalias{Raboonik2024a}). While analytically $\Delta E_\text{tot} = \Delta \mathcal{E}_\text{tot}$, this equality is not guaranteed in a numerical implementation due to inevitable discretization inaccuracies affecting the two quantities in different ways (to be discussed in more detail in Section~\ref{sec:errors}). Note that $\Delta E_\text{tot}$ is computed directly from the solution vector provided by the code's direct numerical simulation scheme without the need for further differentiation, while $\Delta \mathcal{E}_\text{tot}$ follows from the sum of the time integrals of all the component eigenenergies given by Equations~\ref{eq:EEDMNew}, which involve spatial derivatives of the simulated primitive variables.
Defining a net numerical fidelity error $\epsilon_\text{tot}\lrp{\bs{x}, t} = \| \Delta{E}_\text{tot} - \Delta{\mathcal{E}_\text{tot}} \|$, it becomes clear that a perfect numerical EEDM analysis would achieve a value of zero everywhere. Importantly, $\epsilon_\text{tot}$ (or simply Equation~\ref{eq:totalEnergyDecomposition}) is the main constraint for choosing the best numerical scheme to compute the EEDM equations. It was the minimization of $\epsilon_\text{tot}$ that led us to the spatial differentiation and time integration schemes mentioned above. However, these choices may not be fit for simulation codes other than Lare3d. We shall return to this point later in the section.

\subsection{Stratified Gas Model}
\begin{figure}
    \centering
    \includegraphics[width=0.95\linewidth]{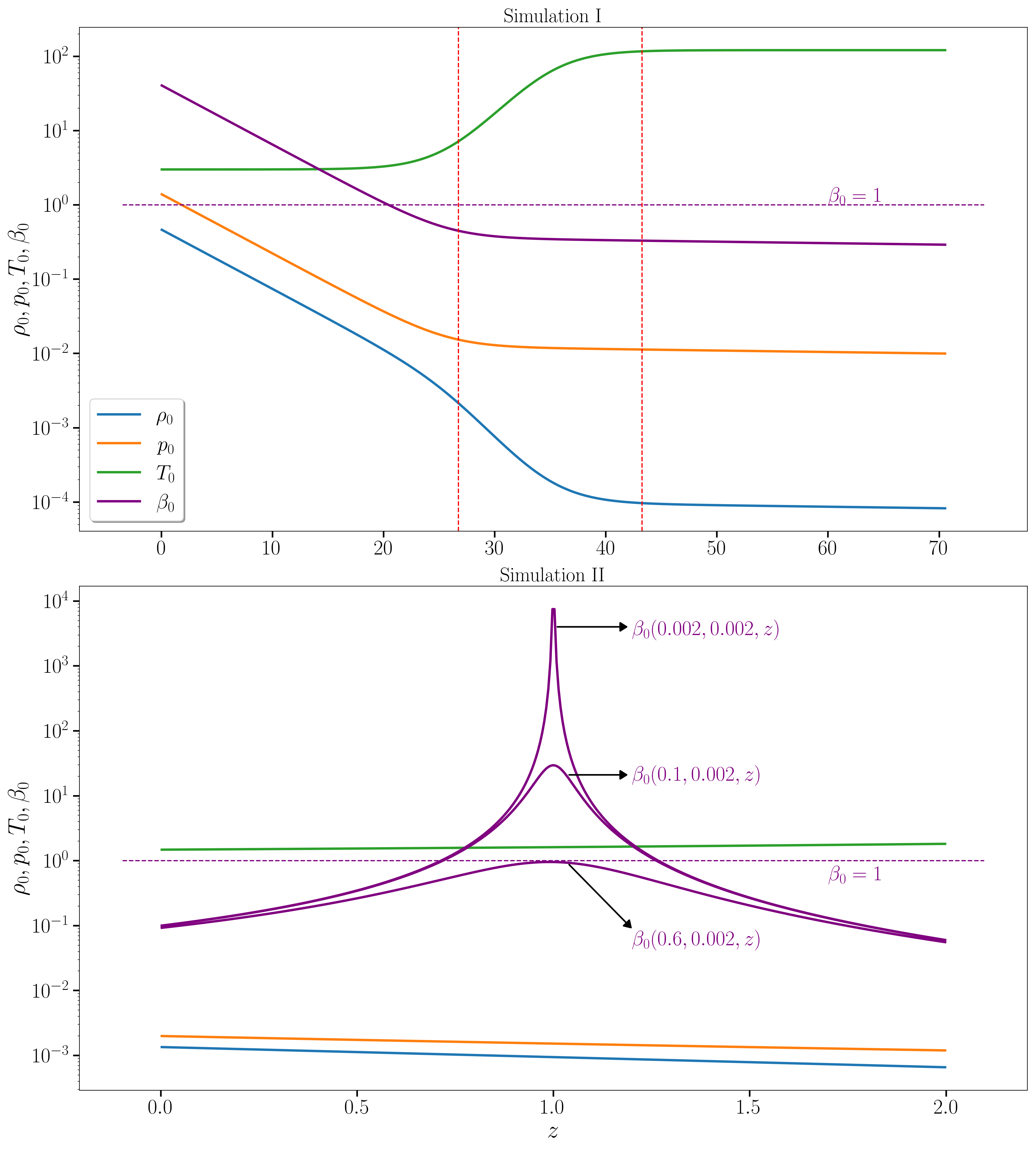}
    \caption{The gravitationally stratified gas models in dimensionless logarithmic scale corresponding to simulations I and II. The vertical dashed red lines in the top panel (Simulation I) delineate the TR. The initial plasma-$\beta$ profile $\beta_0$ in Simulation II is plotted at three horizontal locations corresponding to $(x, y) = ([0, 0.1, 0.6], 0.002)$ as annotated, passing through a magnetic fan plane containing a null point at $(0,0,1)$. The horizontal dashed purple lines are overplotted to mark the $\beta_0 = 1$ locations, whose intersections with $\beta_0$ \emph{roughly} locate the equipartition heights $z_\text{eq}$.}
    \label{fig:gas}
\end{figure}

Conceptually, the models including the background magnetic fields, drivers, and BCs have been carried forward directly from Section 3 of \citetalias{Raboonik2024a}, but now taking on varied strengths and are superposed with 1D gravitationally stratified hydro\-static (HS) ideal gases.
The new gas models associated with the two simulations are adopted from \cite{Tarr2017}, and shown in Figure~\ref{fig:gas} (in nondimensional units). 
Specifically, the background temperature profile is prescribed by \citep[Equation 6 in][]{Tarr2017}
\begin{equation}
    T_0 = T_\text{ph} + \frac{T_\text{cor} - T_\text{ph}}{2}\lrp{\tanh\lrp{\frac{z - z_\text{tr}}{W_\text{tr}}} + 1},
\end{equation}
in which the subscripts ph, cor, and tr stand for photosphere, corona, and transition region (TR), respectively, and $W_\text{tr}$ denotes the width of the TR. 
The background density follows from integrating the hydro\-static equation $\bs{\nabla}p_0 = -\rho_0 \bs{\nabla}\phi$ subject to a constant gravitational force acting downwards along the $z$-axis. That is, $\phi = g z$, where  $g = 274 \tau_N^2 L_N^{-1}$ is the nondimensionalized solar photospheric gravitational acceleration, with $\tau_N$ and $L_N$ denoting the nondimensionalization time and length scales (see Table~\ref{tab}). The pressure (and internal energy) is thence computed via the ideal gas equation of state. The HS model parameters, as well as the Lare3d native parameters associated with both Simulations I and II are listed in Table~\ref{tab}. These parameters, as well as the background magnetic fields (to be discussed), are tuned to position an equipartition layer at a reasonable height inside the simulation box without giving rise to radically high \alf{} speeds causing impractically small time steps.

\begin{table}[]
    \centering
    \begin{tabular}{c||c|c|c}
        & Parameter & Simulation I & Simulation II \\ \hline \hline
        & $L_N$ & $8\times10^{4}$ m & $1.5\times10^{5}$ m \\
        & $\rho_N$ & $3.03\times10^{-4} \text{kg}.\text{m}^{-3}$ & $3.03\times10^{-4} \text{kg}.\text{m}^{-3}$ \\
        & $B_N$ & 0.123 T & 0.1945 T \\
        & $T_N$ & 5778 K & 14445 K \\
        Lare3d & $L_{x\rm{/}y}$ & $-0.1715\leq x,y \leq 0.1715$ & $-1\leq x,y \leq 1$ \\
        & $L_z$ & $0\leq z \leq 70.511$ & $0\leq z \leq 2$ \\
        & $\delta q$ & $4.9\times10^{-3}$ & $5.2\times10^{-3}$ \\
        & $t_\text{end}$ & 38 & 1.5\\
        & $\delta t$ & 0.05 & $3.75\times 10^{-3}$\\
       & visc1 & $10^{-4}$ & $10^{-4}$ \\
       & visc2 & $2\times10^{-5}$ & $2\times10^{-5}$ \\ \hline
        & $\rho_\text{ph}$ & 0.4626 & $1.347\times 10^{-3}$\\
        & $T_\text{ph}$ & 2  & 0.2 \\
       HS model & $T_\text{cor}$ & 120 & 150\\
        & $z_\text{tr}$ & 35 & 15.71 \\
        & $W_\text{tr}$ & 5 & 5 \\
    \end{tabular}
    \caption{Lare3d and HS model parameters used in the two simulations. The numbers without specified physical units are dimensionless. $L$ denotes length, $Q_N$ for any parameter $Q$ denotes the nondimensionalization coefficient, $L_q$ denotes the length of the simulation box in the direction $q\in(x,y,z)$, $\delta q$ and $\delta t$ are respectively the spatial and temporal resolution of the snapshots, and $t_\text{end}$ denotes the duration of the simulations. These parameters are chosen so as to appropriately position an equipartition layer in the simulation box while avoiding extreme \alf{} speeds.}
    \label{tab}
\end{table}

\subsection{Net Branch Eigenenergies}
It is advantageous to define a spatially contracted version of the eigenenergies denoted by $E_b$ called the \emph{net branch eigenenergy}, by summing over all the eigenenergies within each of the MHD branches $b\in [\text{ent}, \text{div}, \text{A}, \text{s}, \text{f}, \text{g}]$. That is, 
\begin{equation}
    E_b = \sum_{q\in\lrp{x,y,z}}\begin{cases}
        E_{b,q} & b\in [\text{ent}, \text{div}, \text{g}] \\
        \sum_{d\in \lrp{-,+}}E_{b,q}^{d} & \text{otherwise,}
    \end{cases}
\end{equation}
and hence $\Delta \mathcal{E}_\text{tot} = \sum_b E_b$ (which analytically also equals $\Delta E_\text{tot}$, but not numerically).
Generally speaking, while $E_b$ is merely an adjunctive quantity and should in no way be taken as a replacement for $E_{m,q}$, it can offer profound insight about the nature of the disturbances, especially in the presence of spatial symmetries. However, bear in mind that the spatial contraction comes at the cost of masking some information, e.g., the reflection of the modes. Thus, in any case, one should always examine and track the individual $E_{m,q}$ components, in particular when analyzing asymmetrical MHD disturbances propagating in all directions\footnote{A conceptual simulation idea for this would be the relaxation of an asymmetrical magnetic twist situated at the center of the box, in which case the $q$- and $d$-directed components would capture the co-directed parts of the propagating energy.}. Here, due to the symmetries of our simulations, for brevity and after careful examinations of both $E_b$ and $E_{m,q}$, we will only discuss the former.

\subsection{Simulation I: Uniform Vertical Magnetic Field in Stratified Atmosphere}\label{sec:simI}
Consider a uniform vertical magnetic field of strength $B_0 = 0.2618$ superimposing the 1D gravitationally stratified HS model containing a smooth corona as shown in the top panel of Figure~\ref{fig:gas}. This results in the initial plasma-$\beta$ profile $\beta_0$ plotted in solid purple, decreasing with height from $\beta = 40.5$ at the bottom (photosphere) to $\beta = 0.29$ at the top. As a consequence, we have $c_{\text{s},z} \approx a_z$ and $c_{\text{s},f} \approx c$ at the bottom, rendering any slow/fast waves magnetically/acoustically dominated, and vice versa at the top. Note that the intersection of this curve with the $\beta_0 = 1$ line only gives an approximation for $z_\text{eq}$. However, the exact equipartition height found by solving $a(z) = c(z)$ for $z$ is $z_\text{eq} = 19.3305$ in this model atmosphere.

Implementing a vortical driver using Equations~\ref{eq:turkmani2004} and fixing the thermodynamic variables to their initial values (Dirichlet BCs) as the bottom BCs, we wish to study the eigenenergy composition of the resultant disturbances. Furthermore, the top and horizontal BCs are set to zero-gradient reflective and periodic, respectively, though we do not analyze the reflected pulse from the top boundary and simply terminate the simulation before high amplitude disturbances hit the top.

\subsubsection{Simulation I: Results}
\begin{figure}
    \centering
    \includegraphics[width=0.95\linewidth]{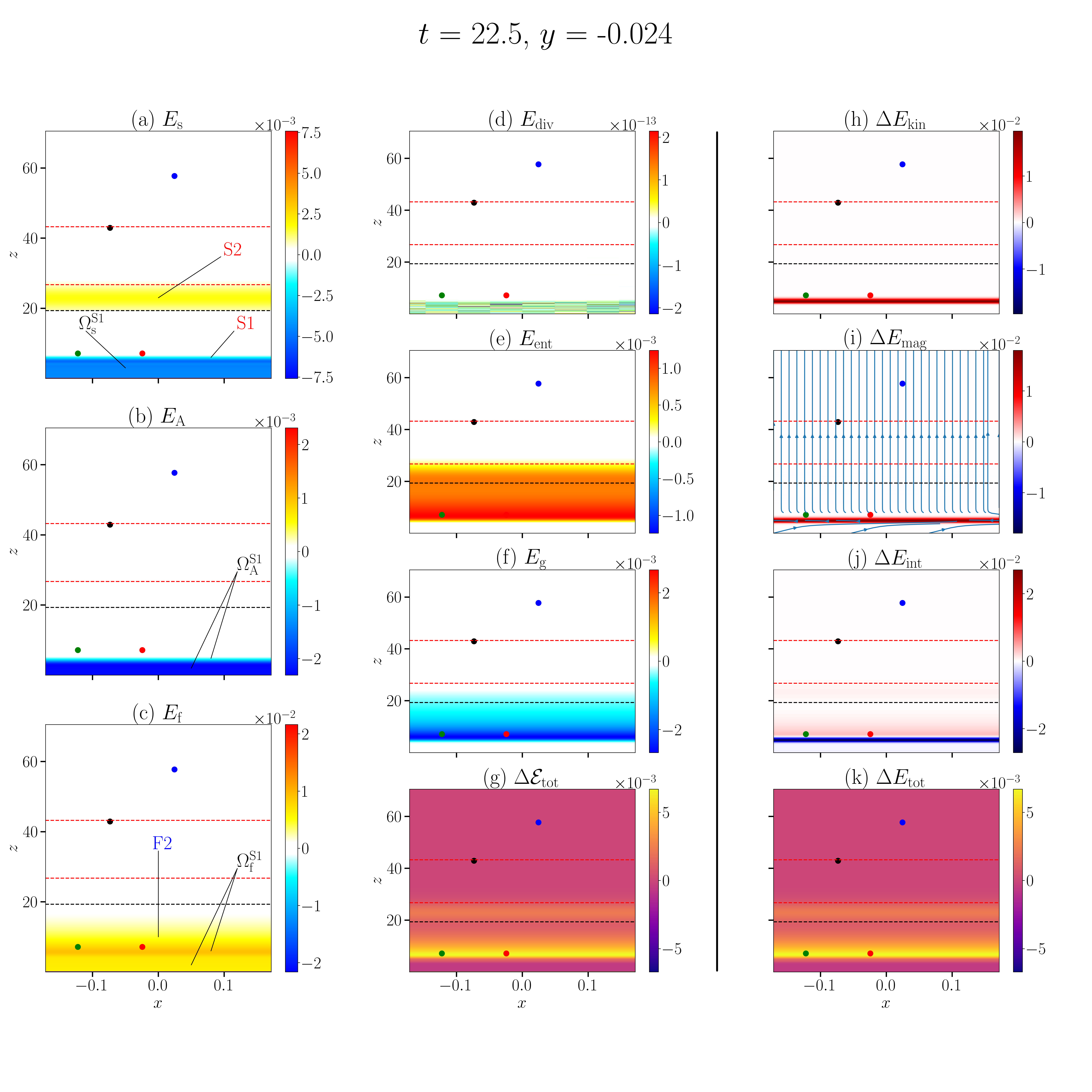}
    \caption{The net branch eigenenergies (a-f), the post-decomposition total energy difference (g), the differences in the kinetic, magnetic, and internal parts of the energy with respect to their initial values (h-j), and the pre-decomposition energy difference (k) associated with Simulation I, plotted as functions of $x$ and $z$ at $y \approx 0$ and $t = 22.5$. The black, red, green, and blue dots, called the wave-tracing markers, locate the position of hypothetical sound, slow, \alf, and fast waves, respectively, by forward time integrating their respective propagation speeds, assuming they have been launched coincidentally with the physical pulse. The labels $\Omega_{b_1}^{b_2}$ denote the non-propagating (PA) natural eigenenergy features, while the non-$\Omega$ labels belong to the propagating natural features. The blue arrows in panel (i) show the magnetic field lines, the horizontal dashed black line marks the equipartition layer, and the horizontal dashed red lines delineate the TR. Note the high amplitude wave in panel (i), warping the magnetic field lines, bearing in mind that the visual skewness of the arrows is exaggerated as a result of the low aspect ratio of the plots.}
    \label{fig:simIEtot2D}
\end{figure}

Figure~\ref{fig:simIEtot2D} depicts the net branch eigenenergies of the pulse at $y \approx 0$ and $t = 22.5$. The panels to the left of the vertical solid black separating line (a-g) are all post-decomposition quantities, while the ones to the right (h-k) correspond to pre-decomposition variables (computed from raw simulation outputs), wherein $\Delta E_\text{kin}$, $\Delta E_\text{mag}$, and $\Delta E_\text{int}$ respectively denote the kinetic, magnetic, and internal energy differences (with respect to their initial amounts). This layout will remain unchanged for all the 2D plots in this paper. The red, green, blue, and black points are hypothetical wave-tracing markers moving along the $z$ direction at $c_{\text{s},z}$, $a_z$, $c_{\text{f},z}$, and $c$, respectively. We will refer to these points simply as the branch markers, whose locations at each time step are determined simply by forward time-integrating their respective speed from $t_0$, assuming initially they all spawn with the pulse. Note that the markers are overplotted merely for illustrative purposes assuring the reader that any \emph{propagating} branch eigenenergy observed does in fact propagate at the right speed, while any \emph{non-propagating} branch eigenenergy may be seen to move at other speeds. 
However, it is important to bear in mind that this apparent movement of non-propagating eigenenergies would be the result of the motion of the specific propagating disturbance that generates them in real time (via Type 2 conversions), not of their own. We shall make the convention of labeling any non-propagating eigenenergy features by $\Omega_{b_1}^{b_2}$, which reads a PA eigenenergy belonging to a branch $b_1$ generated/\emph{powered} by (i.e., Type 2 converted from) a propagating branch $b_2$.
In each panel, the horizontal dashed red lines locate the TR and the horizontal dashed black line marks the equipartition layer $z_\text{eq}$. The magnetic field lines are overplotted in blue in panel (i). It is seen that the pulse is composed of all the branch eigenenergies, except for $E_\text{div}$ in panel (d), where the features are noise-like and of (numerically) ignorable magnitude. We shall parse the results as follows.

First, focus on panels (a) and (c), and note the location of the fast marker (blue point), which has traveled well above the TR. The absence of any fast branch eigenenergy features there in panel (c) suggests that either the driver did not produce any fast wave, or if it did, then something must have happened to the fast wave somewhere along the way, impeding it from reaching these height. Also observe the slow marker (red point), sitting well below the upper feature labeled S2 in panel (a) belonging to the slow branch. Since any driven slow wave could not have traveled that far, the S2 feature must have been converted from a faster traveling wave. Combining these observations, we infer that an initial upward-propagating fast pulse, call it F1, must have been generated at the bottom, fully (Type 1) \emph{converted} to a propagating slow wave S2 at and in the close proximity of $z_\text{eq}$, and from there been denied transmission through the barrier of the TR as expected--while the purely magnetic nature of \alf{} waves allows them to penetrate through TRs, the steep thermodynamical gradients accompanying TRs (which typically involve an exponential decline in $\rho$) can heavily impact magneto\-acoustic (i.e., slow and fast) waves.

\begin{figure}
    \centering
    \includegraphics[width=0.95\linewidth]{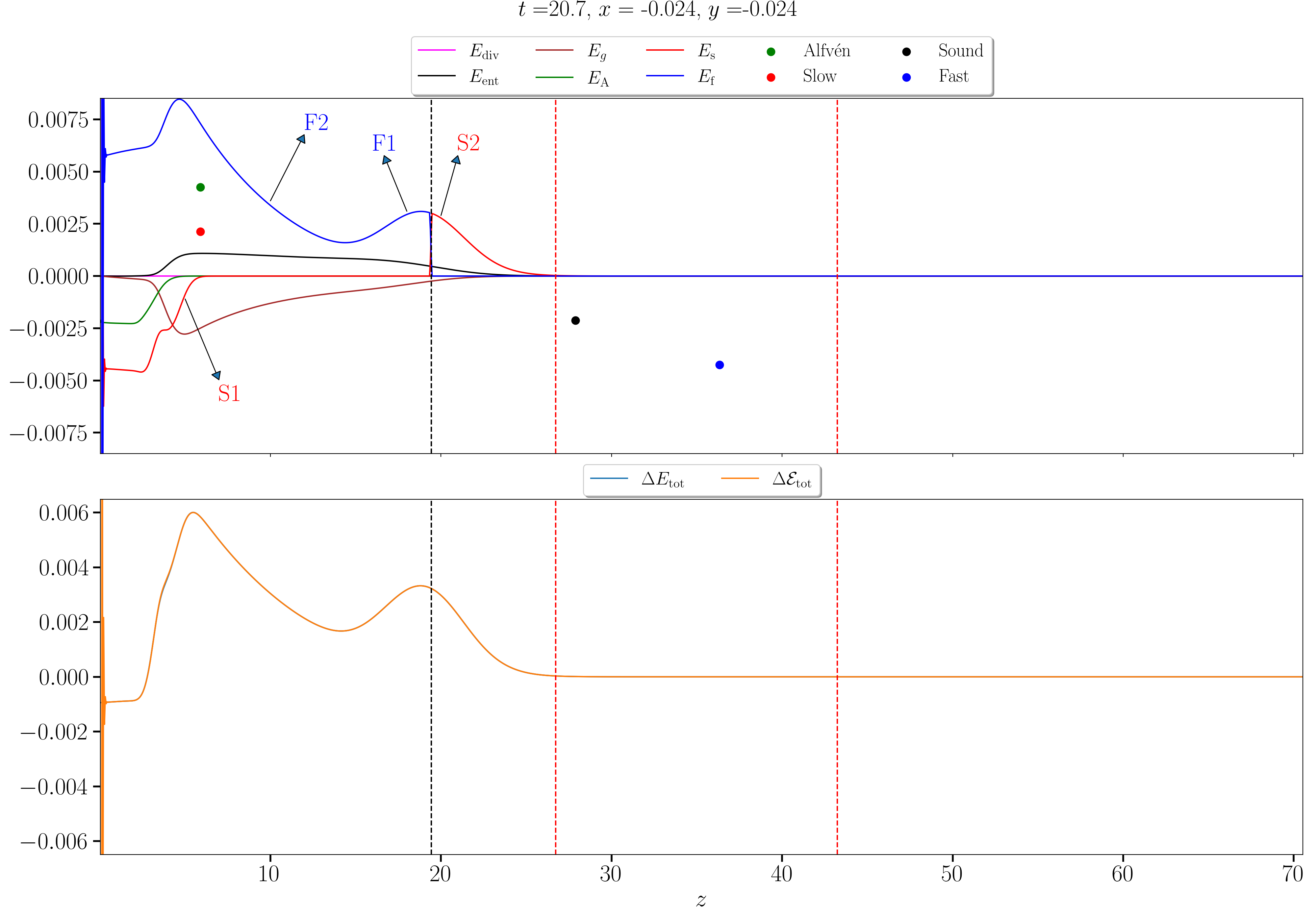}
    \caption{\emph{Top:} The net branch eigenenergies of Simulation I as functions of $z$ plotted at $x = y \approx 0$ and $t = 20.7$. \emph{Bottom:} The pre- and post-decomposition total energy differences associated with the top panel. The colorful dots are the wave-tracing markers. The dashed black and red lines represent the equipartition layer and the two ends of the TR, respectively. The arrows mark four propagating MHD waves, namely, an initial slow (S1) and fast (F1) wave generated by the driver, followed by a mode-converted slow (S2) and fast (F2) wave.
    Notice the high accuracy of the method as signaled by the near-perfect agreement between $\Delta E_\text{tot}$ and $\Delta \mathcal{E}_\text{tot}$ in the bottom panel.}
    \label{fig:simIEtot1D}
\end{figure}

\begin{figure}
    \centering
    \includegraphics[width=0.95\linewidth]{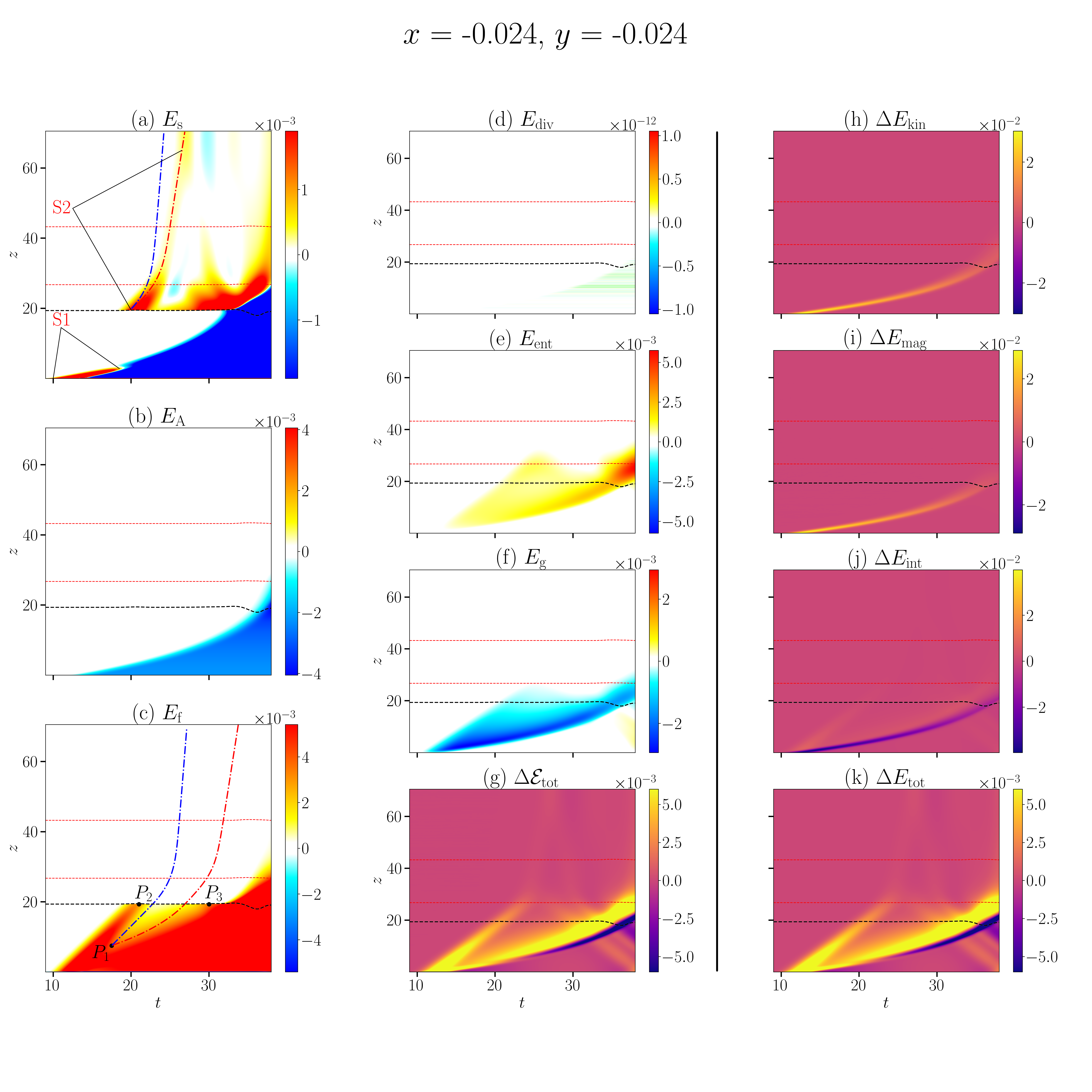}
    \caption{Time-distance ($t$-$z$) plots of the branch eigenenergies associated with Simulation I. The dash-dotted red and blue curves in panels (a) and (c) are respectively hypothetical slow and fast wave trajectories seeded (launched) at their lower end, provided as a reference to compare with the slope (propagation speed) of the slow and fast branches. The slow (a) and fast (c) branches are over-saturated to detect any order-of-magnitude-smaller amplitudes in the vicinity of $z_\text{eq}$. As can be seen, the entire fast energy (traveling upward at $c_{\text{f},z}$; panel c) is first mode-converted into slow energy (traveling upward at $c_{\text{s},z}$; panel a) at $z_\text{eq}$, and from there heavily geometrically attenuated by the TR (dotted red horizontal lines). The points $P_1$ and $P_2$ roughly locate the S1$\overset{1}{\rightarrow}$F2, F1$\overset{1}{\rightarrow}$S2 conversion zones, respectively, and $P_3$ marks where S1 hits $z_\text{eq}$.}
    \label{fig:simIEtotPixel}
\end{figure}

This sharp \modeconvert{1}{F1}{S2} conversion followed by the geometric wave attenuation caused by the TR is also clearly seen in the 1D plot shown in the top panel of Figure~\ref{fig:simIEtot1D} taken at $x = y \approx 0$ and $t = 20.7$. In there, note how the fast (blue) and slow (red) curves tesselate neatly (and quite sharply) at $z_\text{eq}$, as if they belong to the same function. However, this is not simply a ``change of subscripts'' from f to s, but rather a physical conversion of energy from one eigenmode to another, after which (the mode-converted wave) S2 starts propagating at $c_{\text{s},z}$ (see the dash-dotted red curve in panel (a) of Figure~\ref{fig:simIEtotPixel}; we shall come back to this shortly). Mathematically, the variable acting as a switch in charge of slow$\leftrightarrow$fast conversions is $\alpha_{\text{s/f},q}$ (see its definition right below Equations~\ref{eq:EEDMNew}), which can vary dramatically with the plasma-$\beta$ and behave as a smooth Heaviside function around $z_\text{eq}$. In fact, this is the variable that determines whether a slow/fast wave is more magnetically dominated or acoustically \citep[also see][]{roe1996}.

Let us turn our attention back to Figure~\ref{fig:simIEtot2D} and zoom in on any non-propagating eigenenergies. As discussed in Section~\ref{sec:modeConversion}, the passage of nonlinear disturbances through the background medium (described by $E_0$), in our case F1, restructures the disturbed regions according to the ideal-MHD equations. In the language of the eigenmodes, this reconfiguration is done through Type 2 mode conversions into non-propagating (zero-frequency) variations in the eigenenergies, such that the net effect describes the new state. 
In \citetalias{Raboonik2024b}, we investigated this property in detail, and ascertained that while the natural eigenmodes have a dual nature and describe both propagating and (degenerate) non-propagating disturbances, the PA eigenmodes, i.e., div, ent, and now g are strictly non-propagating everywhere. This means that the PA modes would likely always contribute to any background changes, and hence the presence of the positive and negative features in panels (e) and (f), with only a (noise-like) smattering of $E_\text{div}$ in panel (d) implying that the reconfiguration did not need this eigenmode.
Moreover, the fact that none of the PA modes have made it beyond the base of the TR (lower dashed red line) indicates that the initial F1 pulse has almost perfectly been halted by the TR, leaving the state above the TR unaltered.

What about the features S1, $\Omega_{\rm s}^{\rm S1}$, $\Omega_{\rm A}^{\rm S1}$, and $\Omega_{\rm f}^{\rm S1}$ in panels (a), (b), and (c), sitting right at and below the slow/\alf{} markers in Figure~\ref{fig:simIEtot2D}, seemingly moving at $c_{\text{s},z}$ ($\approx a_z$ due to the locally high plasma-$\beta$)?
Do these belong to a slow wave, or an \alf{} wave, or both? What is certain is that they definitely cannot come from a fast wave due to the propagation speed $c_{\text{f},z} > c_{\text{s},z}, a_z$. But to answer these questions precisely, it is important to first take notice of the only feature in panel (i) belonging to $\Delta E_\text{mag}$. Note how it has \emph{warped} the background magnetic lines, signaling a highly nonlinear wave. Note further that it is coincident with similar features in panels (h) and (j) belonging respectively to $\Delta E_\text{kin}$ and $\Delta E_\text{int}$. This coincidence implies a part magnetic and part acoustic nature of this high amplitude wave, ruling out the 
\alf{} mode (due to the purely incompressive magnetic nature of \alf{} waves). Thus, we identify this wave, whose front appears as the upper thin stripe in panel (a) labeled S1, as a pure high amplitude propagating slow wave named after its label.

This leaves us with two other questions: if S1 is in fact a slow wave, then why is it negative in amplitude, and why are there other non-slow eigenenergy features adjacent (from above and below) to it? The answer to both of these questions yet again lies in nonlinear mode conversion (see Section~\ref{sec:modeConversion}) and is as follows. Firstly, the passage of the high amplitude propagating slow wave S1 modifies the background plasma (described by $E_0$), leading to Type 2 conversions of energy into non-propagating eigenenergies. This gives rise to the $\Omega_{\rm s}^{\rm S1}$, $\Omega_{\rm A}^{\rm S1}$, and $\Omega_{\rm f}^{\rm S1}$ features in panels (a)\footnote{Interestingly, the new state created by the slow wave is in part expressed in terms of some non-propagating slow eigenenergy.}, (b), and (c).
Another piece of evidence that S1 is indeed the origin of said non-propagating features is the fact that it is simply the leading feature running ahead of all the other branches below the slow marker. This is clearly evident in the top panel of  Figure~\ref{fig:simIEtot1D}, wherein the tip of the slow (red) curve segment to the left of the slow marker is ahead of the other nearby features. 
Secondly, in addition to the Type 2 mode conversions, S1 is also seen to undergo a Type 1 conversion into some (additional) propagating fast wave (F2). Thus, it is these conversions that leaves S1 with a negative eigenenergy.

Note that when examining static single-snapshot plots such as Figures~\ref{fig:simIEtot2D} and \ref{fig:simIEtot1D}, aside from the chain of reasoning discussed above based on the observed locations of the features and their comparison with the branch markers, there is no direct way of distinguishing between the propagating and non-propagating/PA features. However, time-dependent visualizations of the eigenenergies such as time-distance plots (to be discussed below) and animations (attached to the paper) can offer a detailed view into the propagating or PA essence of the features. In any case, and as touched on before, any local eigenenergy feature of a certain branch $b_1$ that seems to move at the (characteristic) speed of some other (natural) branch $b_2$ should be taken as a clear sign that the $b_1$ feature is non-propagating. This means that $b_1$ was created through Type 2 mode conversion as a response to any background modifications caused by $b_2$.

To further corroborate our analysis so far, we shall examine the full time evolution of the branches at $x = y \approx 0$ as shown in the time-distance plots of Figure~\ref{fig:simIEtotPixel}. 
The colormaps in panels (a) and (c) are over-saturated by an order of magnitude to reveal small amplitude slow and fast features.
In these panels, the dash-dotted red and blue curves represent vertical-only space-time \emph{wave trajectories} (WTs) belonging to hypothetical slow and fast waves, respectively. These WTs are simply the counterparts of the branch markers, derived by numerically solving $dz = u dt$ from a given seeding point ($t_\text{min}, z_\text{min}$) at fixed $x$- and $y$-coordinates, where $u \in [c_{\text{s},z},a_z,c_{\text{f},z}]$ is the vertical propagation speed. Therefore, the slope of the WTs provides a good visual benchmark for the vertical speed of propagation of the eigenenergy features.

Notice the feature labeled S1 at the bottom left of panel (a) that extends over $10 \lessapprox t \lessapprox 18$ and $0 \lessapprox z \lessapprox 4$. This is indeed the slow pulse S1 before undergoing the above mentioned Type 1 and 2 conversions that follow from $t \approx 18$, when it starts declining in amplitude until it fully falls below zero at $t \approx 19$. This coincides with the triangular feature in panel (c), whose vertices are marked by $P_1$, $P_2$, and $P_3$, indicating the Type 1 conversion that generates F2 (compare the slope of the features with the overplotted WTs). 
Here, $P_1$ and $P_2$ respectively mark the locations of the \modeconvert{1}{S1}{F2} and \modeconvert{1}{F1}{S2} conversions, and $P_3$ locates the intersection of S1 with $z_\text{eq}$. 
The upper lobe of the features below $z_\text{eq}$ in panel (a) (comprised of the S1 feature and its negatively valued extension) traces out the trajectory of the S1 wave-front, and everything below this lobe in panels (a), (b), and (c) are the PA features $\Omega_{\rm s}^{\rm S1}$, $\Omega_{\rm A}^{\rm S1}$, and $\Omega_{\rm f}^{\rm S1}$ described in Figure~\ref{fig:simIEtot2D}. 
Moreover, panels (h), (i), and (j) indicate that S1 is characterized by a drop in the background internal energy, and gains in the magnetic and kinetic energies.
In these observations, note that both of the Type 1 conversions resulted in propagating waves whose propagation speeds are indeed different from their parent wave. On the other hand, the Type 2 conversion from S1 created both slow and non-slow non-propagating features, whose fronts co-move with the S1 front at the speed of $c_{\text{s},z}$.

To summarize Simulation I, the driver generates a highly nonlinear slow wave, and a fast wave of relatively smaller amplitude, namely, S1 and F1, respectively (see Figure~\ref{fig:simIEtot1D}). As they propagate through the plasma, both of these initial waves undergo both Type 1 and 2 mode conversions at different locations. The type 1 conversions leading to secondary propagating slow and fast waves are \modeconvert{1}{S1}{F2} and \modeconvert{1}{F1}{S2}, which respectively take place at $z \approx 4$ and $z = z_\text{eq}$. The Type 2 conversions as a result of background modifications appear in Figure~\ref{fig:simIEtot2D} as the $\Omega_{\rm s}^{\rm S1}$, $\Omega_{\rm A}^{\rm S1}$, and $\Omega_{\rm f}^{\rm S1}$ features below the slow marker in panels (a), (b), and (c) for S1, and the features above this maker in panels (e) and (f) for F1. 

Finally, note the near perfect agreement between $\Delta E_\text{tot}$ and  $\Delta \mathcal{E}_\text{tot}$ in all three of the figures discussed above, where computations of $\epsilon_\text{tot}$ show a local maximum of 3\% error. 

\subsection{Simulation II: 3D Structured Magnetic Field with a Null}
For a more involved illustration of the numerical applications of the EEDM, let us consider the 3D torsional wave-null interaction simulation setup of  \citetalias{Raboonik2024a} (Section 4 therein; originally based upon \cite{Galsgaard2003}), but now superimposed with the stratified gas model shown in the bottom panel of Figure~\ref{fig:gas}. 
The background magnetic field follows from (the Equation 15 in \citetalias{Raboonik2024a})
\begin{equation}\label{eq:galsgaardfield}
    \bs{B}_0 = B_0 \lrp{x, y, -2\lrp{z - 1}},
\end{equation}
where the null is located at $(0,0,1)$. In this simulation setup, we set $B_0 = 0.1$ and the gas model parameters are found  in Table~\ref{tab}. This results in an initially ellipsoidal equipartition layer whose boundaries are at 
$-0.549<x_\text{eq}(y=0,z=1),y_\text{eq}(x=0,z=1)<0.549$ and $0.71<z_\text{eq}(x=y=0)<1.268$, and the local plasma-$\beta$ profiles in the bottom panel of Figure~\ref{fig:gas} shown in purple plotted at $y = 0.002$ and three different locations along the $x$-axis.
Thus, we are dealing with a relatively low-$\beta$ plasma outside of the ellipsoid, indicating that any slow/fast wave in this region would be acoustically/magnetically dominated with $c_\text{f} \approx a$, while the situation is reversed within the ellipsoid, where $c_\text{s} \approx a$. 
To torsionally excite propagating nonlinear waves, the base of the magnetic spine at $z = 0$ is twisted by employing the 2D vortical velocity and magnetic fields given by Equation~\ref{eq:galsDriver} (same as the Equations 16 and 17 of \citetalias{Raboonik2024a}) at high amplitudes and enforcing Dirichlet BCs on the gas model at the bottom. The other BCs for the remaining five faces are all set to zero-gradient reflective.

\subsubsection{Simulation II: Results}
\begin{figure}
    \centering
    \includegraphics[width=0.95\linewidth]{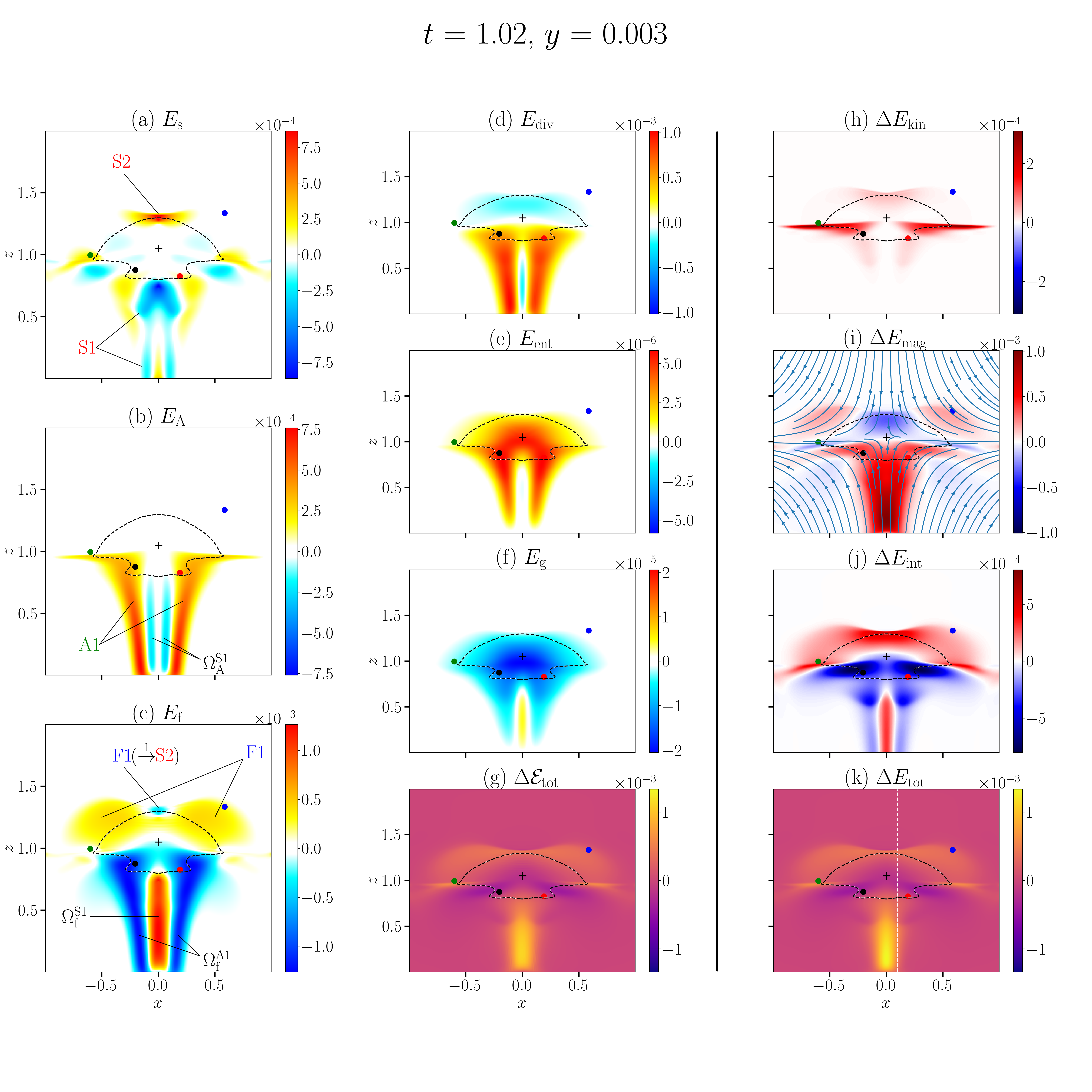}
    \caption{The net branch eigenenergies, energy difference components, and the pre- and post-decomposition total energy differences of Simulation II plotted at $y \approx 0$ and $t = 1.02$, presented in the same fashion as Figure~\ref{fig:simIEtot2D}. The near-central plus sign marks the magnetic null, and the dashed black contour surrounding it is the equipartition layer, which has evolved to this shape from an initial ellipsoid, indicating the highly nonlinear nature of the propagating waves in this simulation. The vertical dashed white line in panel (k) marks the data-slicing location for the subsequent spatially 1D figures.}
    \label{fig:simIIEtot2D}
\end{figure}

\begin{figure}
    \centering
    \includegraphics[width=0.95\linewidth]{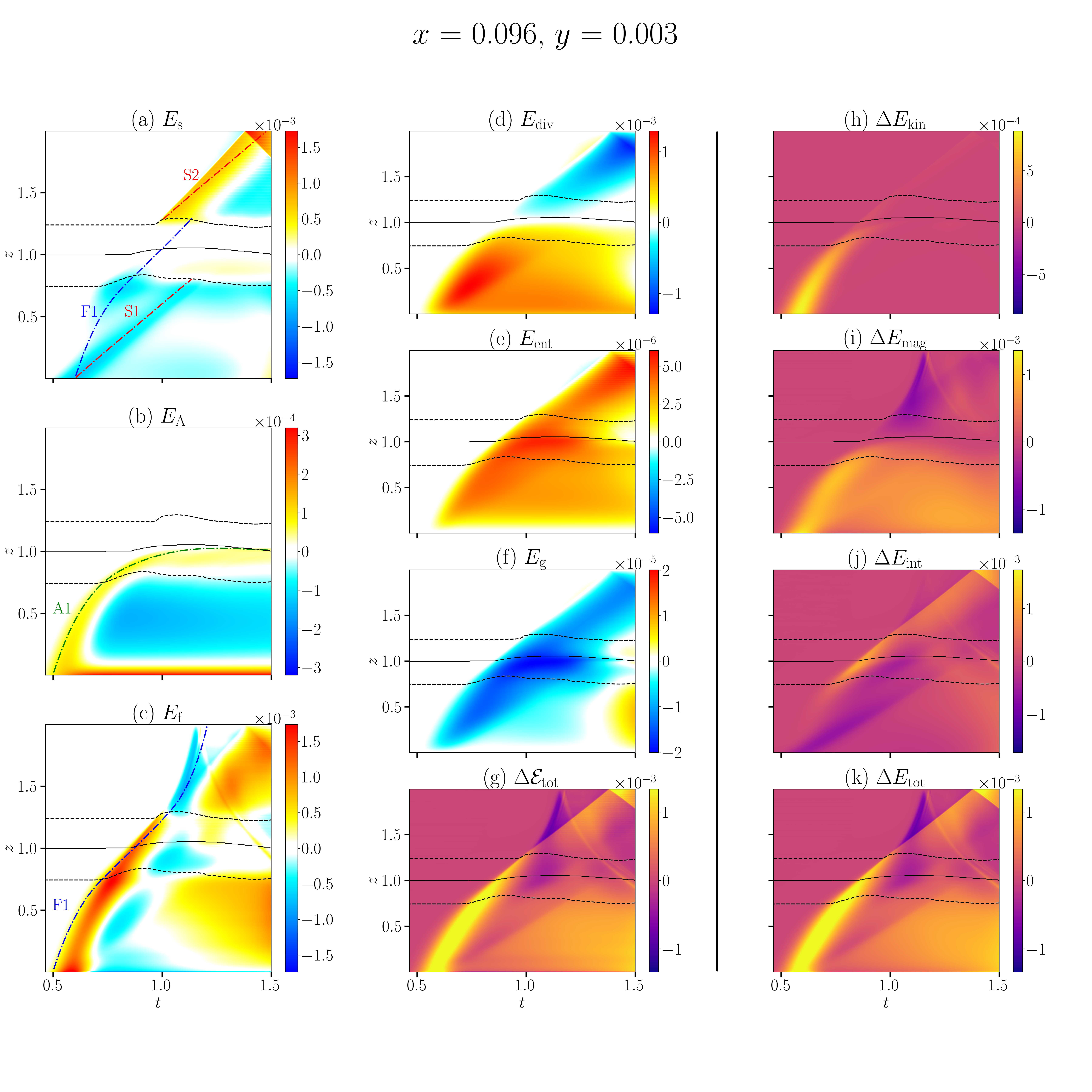}
    \caption{Time-distance plots of the net branch eigenenergies of Simulation II at $(x,y) \approx (0.1, 0)$, following the same layout as Figure~\ref{fig:simIEtotPixel}. The horizontal solid and dashed black curves respectively mark the location of the magnetic null and the top and bottom of the EC.
    In addition to the slow and fast WTs respectively shown in the dash-dotted red and blue curves, we have overplotted an \alf{} WT in dash-dotted green in panel (b). Here S1, F1, and A1 represent propagating nonlinear waves generated by the driver, and S2 labels a mode-converted propagating slow wave.}\label{fig:simIIEtotPixel}
\end{figure}

The wave analysis can be done in a similar fashion to Simulation I. Figure~\ref{fig:simIIEtot2D} depicts the 2D $x$-$z$ net branch eigenenergies of Simulation II, wherein the black dashed loop (resembling a mushroom) delineates the equipartition contour (EC), heavily modified from the initial ellipsoidal shape by the nonlinear waves, the (near-central) black plus sign locates the magnetic null, and the dashed white line in panel (k) is where the data will be sliced ($x_\text{slice} = 0.096$) for the subsequent (time-distance and 1D) plots (see Figure~\ref{fig:simIIEtotPixel} and Figure~\ref{fig:simIIEtot1D}). 

Examining panel (a) of this figure, we immediately identify a similar occurrence to that of Simulation I: the emergence of $E_\text{s}$ at the top of the EC labeled S2. Since any slow wave generated at the bottom boundary could not have traveled this far, this must be the result of a \modeconvert{1}{fast}{slow} conversion. Let us refer to the (parent) fast and (mode-converted) slow waves as F1 and S2, respectively. Additionally, notice the S1 feature extending from the bottom to the base of the EC. This suggests a highly nonlinear slow wave S1, which is in part responsible for significantly modifying the background plasma and uplifting the lowermost lobe of the EC, reshaping it into the bottom flat segment. The fact that S1 has ``squished'' the EC and elevated it upwards is a clear indication that it must have been denied transmission/entry into the high-$\beta$ region inside the EC.
Moreover, the negative eigenenergy corresponding to S1 implies it must have undergone some mode-conversion.

In panel (b), note the complete blockade of all the \alf{} features by the magnetic fan, on which $a_z = 0$. This is the expected behavior of any true \alf{} wave \citep{Raboonik2024a,Galsgaard2003}. Note further the alignment of the top of the A1 features at the fan with the \alf{} marker and the sides of the EC ($z \approx 1$). This means that A1 is a high-amplitude \alf{} wave also generated by the driver, propagated along the field lines towards the fan while reconfiguring the background plasma in real time giving rise to the wings of the EC, and finally halted by the fan. The two central spikes labeled $\Omega_{\rm A}^{\rm S1}$ are non-propagating \alf{} features due to S1.

Having identified F1, S1, S2, and A1, it becomes relatively easier to parse the $E_\text{f}$ features in panel (c). Starting from the top, the negative feature labeled F1($\overset{1}{\rightarrow}$S2) is the fast wave F1 undergoing the Type 1 conversion into S2, the positive features F1 are the  \emph{transmitted} parts of F1 propagating out of the EC, and the $\Omega_{\rm f}^{\rm A1}$ and $\Omega_{\rm f}^{\rm S1}$ features are respectively the result of \modeconvert{2}{A1}{fast} and \modeconvert{2}{S1}{fast} conversions into non-propagating fast eigenenergies.
Finally, panels (d), (e), and (f) detect the three inherently non-propagating PA eigenenergies generated as a result of the background modifications induced by the propagating waves.

Figure~\ref{fig:simIIEtotPixel} shows the time-distance plots of the net branch eigenenergies at $x_\text{slice} = 0.096$ and $y_\text{slice} = 0.003$ (along the dashed white line in panel (k) of Figure~\ref{fig:simIEtot2D}). The dashed black horizontal curves represent the two ends of the EC at $(x_\text{slice},y_\text{slice})$, and the middle solid black horizontal curve marks the location of the null/fan. 
In panels (a)-(c), we have overplotted some slow, fast, and now \alf{} WTs in dash-dotted red, blue, and green curves, respectively. 
To remind the reader, these are $t$-$z$ trajectories of hypothetical natural propagating waves obtained by forward time integrating $dz = u dt$ from some seeding (or initial) point ($t_\text{min}, z_\text{min}$), where $u$ is the $z$-component of any of the three characteristic speeds at $(x_\text{slice},y_\text{slice})$.
To use their local slopes as a benchmark for comparison with the eigenenergies, their seeding points are chosen to roughly collocate with the base of the S1, S2, F1, and A1 features. Given the 2.5D nature of the simulation, this procedure closely approximates the vertical energy propagation of each mode, as can be seen by comparing the local slope of each curve with the nonlinear eigenenergy features. The correspondence is very good in the lower portions of the plots, where the vertical components of the propagation of the azimuthally symmetric pulses closely coincide with our calculation, while the departures in the upper portion of the plots stem from projections of the now-obliquely propagating modes across our chosen fixed trajectory.  The behavior of each mode is still abundantly clear in these upper regions.

Note the perfect halting of A1 by the fan in panel (b), the near-perfect halting of S1 by the base of the EC in panel (a), and the partial transmission of F1 through into the EC and the subsequent \modeconvert{1}{F1}{S2} conversion upon leaving the upper EC. This conversion is in essence of the same origin as the \modeconvert{1}{F1}{S2} conversion of Simulation I, in that a fast wave converts into a slow wave in going from a high-$\beta$ region into a low-$\beta$ one. Notice that the $E_\text{s}$ associated with S1 in panel (a) appears as negative throughout the simulation. This indicates that S1 must have been partially converted into other propagating modes in the ghost cells before reaching the main domain.
Although nothing precise can be inferred about this conversion due to the lack of data in the subsurface regions, comparing the amplitudes in panels (a) and (b) may be an indication that A1 was born out of S1. 
Nevertheless, the salient point here is that the eigenenergy identification is precise regardless of such conversions.

As for panels (d), (e), and (f), the presence of some small amounts of $E_\text{ent}$ and $E_\text{g}$, and order-of-magnitude amplitudes of $E_\text{div}$ indicates heavy modifications to the background plasma, especially to the magnetic field. Furthermore, the void of $E_\text{A}$ above the fan in panel (b) implies that not only was A1 blocked by the fan, but no further \alf{} waves could be generated through mode-conversions in this simulation setup. In other words, the incompressible torsional behavior of the disturbances was filtered out by the fan. 
Lastly, note that even though the difference between the magnitudes of the characteristic speeds is not significantly large in this simulation, the EEDM precisely and uniquely detects all of the eigenenergies generated by the driver, without any source of cross-eigenenergy contamination.

In the following section, we will take the EEDM analysis of Simulation II presented here as an example in order to discuss the potential sources of discretization inaccuracies arising in numerical implementations of the method.

\subsection{Potential Sources of Numerical Errors}\label{sec:errors}
\begin{figure}
    \centering
    \includegraphics[width=0.9\linewidth]{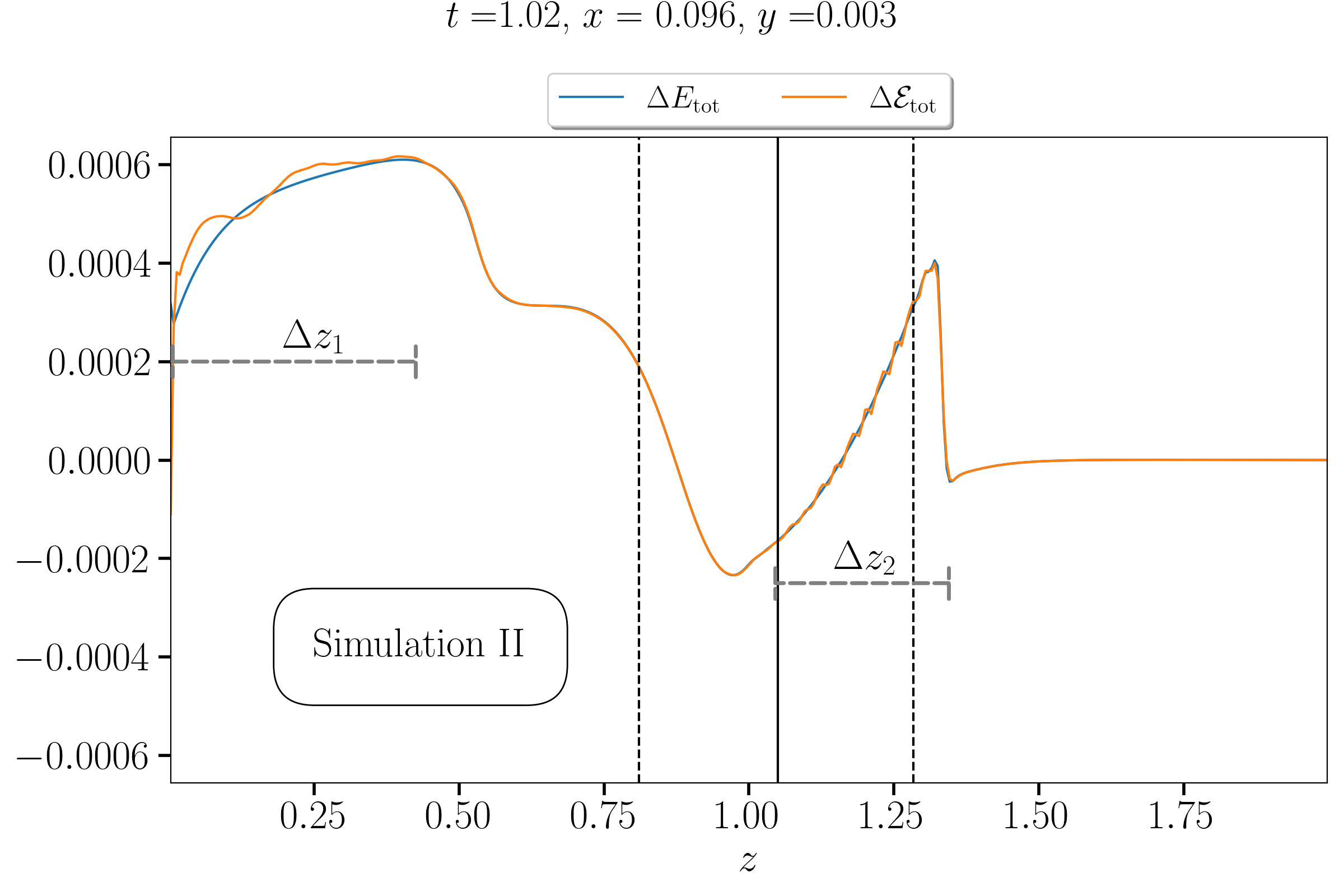}
    \caption{The pre- and post-decomposition total energy differences associated with Simulation II at the location of the vertical dashed white line in panel (k) of Figure~\ref{fig:simIIEtot2D}.}
    \label{fig:simIIEtot1D}
\end{figure}

As mentioned before, while the EEDM itself is analytically exact, its numerical implementation may lead to inaccuracies, which manifest as disagreements between the pre- and post-decomposition total energy differences, namely, $\Delta E_\text{tot}$ and $\Delta \mathcal{E}_\text{tot}$.
Let us demonstrate this by taking a close look at these quantities in  Simulation II, as shown in the 1D plot of Figure~\ref{fig:simIIEtot1D} taken at $(x_\text{slice}, y_\text{slice})$.
Notice the patent mismatch between $\Delta E_\text{tot}$ and $\Delta \mathcal{E}_\text{tot}$ over the two intervals labeled $\Delta z_1$ and $\Delta z_2$. 
This provides us with an excellent example as each of the intervals pertains to a different source of numerical errors.

The errors over $\Delta z_1$ originate from the dynamic bottom BCs. This is because the driver (given by Equations~\ref{eq:galsDriver} and enforcing the bottom Dirichlet BCs on the gas) set up in the ghost cells is not consistent with a true ideal MHD evolution, since there are no analytical nonlinear solutions describing propagating torsional ideal-MHD waves \citep{Raboonik2024a}.
Of course, within the physical domain the variables are consistent with the MHD equations, and this creates a mismatch in the vicinity of the bottom boundary. This mismatch (jump) results in some numerical dissipation (resistivity and viscosity) rendering the generated solutions along $\Delta z_1$ non-ideal-MHD in nature. Since the current EEDM is built upon the ideal-MHD equations, a mismatch is observed between $\Delta E_\text{tot}$ and $\Delta \mathcal{E}_\text{tot}$ (see Section 4 of \citetalias{Raboonik2024b}).

While such numerical errors may occur anywhere non-ideal effects become significant in a simulation, the errors across $\Delta z_2$ are predominantly due to the numerical differentiation and integration schemes used to evaluate the EEDM equations, which may be heavily influenced by the spatial and temporal resolution of the solutions. The jagged patterns appearing in $\Delta \mathcal{E}_\text{tot}(z\in \Delta z_2)$ are a sign of round-off errors associated with the numerical evaluation of the spatial derivatives in Equations~\ref{eq:EEDMNew} accumulated over time (as a result of the time integration of Equation~\ref{eq:eigenenergy}). The incremental round-off errors are due to the steep change in the total energy near the right end of $\Delta z_2$. Thus, using higher spatial resolution in the simulation can be a potential easy fix for this problem. Nevertheless, observe the high precision of the EEDM in capturing even poorly resolved shock behaviors, underscoring the strength and robustness of the method.

To summarize, there are two main sources of discretization errors which may arise (either individually or in tandem) in numerical applications of the EEDM, resulting in $\Delta E_\text{tot} \neq \Delta \mathcal{E}_\text{tot}$.
First, any departure in the simulation code from the strict ideal-MHD regime would generate non-ideal behaviors in $\Delta E_\text{tot}$, which fall out of the current scope of the EEDM as described by Equations~\ref{eq:EEDMNew}. Therefore, this roots entirely in the simulation itself, including the simulation code and the spatial resolution of the snapshots.
Second, the numerical schemes chosen to compute Equations~\ref{eq:EEDMNew} and \ref{eq:eigenenergy}, as well as the temporal resolution (cadence) of the solutions can produce further errors in $\Delta \mathcal{E}_\text{tot}$. Thus, this is entirely independent of the simulation code, but rather roots in the EEDM code.

\section{Conclusion}\label{sec:conc}
This paper presents an exact mathematical method for the unique eigenmode decomposition of the energy associated with general nonlinear gravitational ideal-MHD disturbances. Additionally, it addresses the problem of nonlinear MHD mode conversion, and sheds more light on the true nature of the propagating (natural) and non-propagating (PA) modes. Thus, it extends and completes the results of \citetalias{Raboonik2024a} and \citetalias{Raboonik2024b}. Furthermore, Equations~\ref{eq:EEDMNew} and \ref{eq:eigenenergy}, which form the core of the EEDM, offer an alternative view of the gravitational ideal-MHD equations through the evolution of the linearly independent eigenenergies.
Importantly, although the derivations of these equations were demonstrated for the set of gravitational ideal-MHD PDEs, nevertheless, the same guidelines can be applied to any system that obeys general inhomogeneous quasi-linear PDEs under the constraint of a global conservation law.

As was seen in Section \ref{sec:modeConversion}, one powerful capability of the EEDM is the tracking and measuring of nonlinear MHD mode conversions. This is a natural consequence of the definition of the eigenenergies through Equation~\ref{eq:eigenenergy}, allowing them to store negative values, which in turn signal two categories of mode conversion, namely, Type 1 and Type 2. Type 1 conversions allow for a propagating eigenenergy to undergo transformation (partially or entirely) into one or more propagating eigenenergies. Type 2 conversions, on the other hand, describe scenarios involving a propagating eigenenergy (partially or entirely) transforming into one or more non-propagating PA eigenenergies (see Table~\ref{tab:modes} for the relevant properties and origins of the propagating and PA modes).

Consolidating the results of \citetalias{Raboonik2024b} with this paper, it was inferred that the PA eigenenergies are generated as a response to modifications inflicted upon the background plasma as a result of the passage of nonlinear propagating waves. In other words, propagating disturbances evolve the background state into new states, which are then described by the PA eigenenergies. This led us to classifying the eigenmodes into two categories of (\emph{i}) natural propagating modes, subsuming the A, s, and f eigenmodes ($m = [3..8]$), and (\emph{ii}) PA modes, which include the inherently non-propagating ent, div, and g modes, as well as non-propagating A, s, and f eigenmodes. Visually, any non-propagating eigenenergy feature is usually seen to \emph{tailgate} the natural propagating nonlinear wave that generates it. In other words, such a nonlinear wave would always be the leading eigenenergy feature, leaving behind non-propagating features of any mode in its wake. Generally, the natural propagating eigenenergies move at their respective characteristic speeds, while non-propagating eigenenergies move at the characteristic speed of their propagating nonlinear progenitor. For example, a slow eigenenergy feature observed to move at the fast speed means the feature is non-propagating/PA. Of course, in the presence of large bulk motions, one has to also take into account the local direction and magnitude of the velocity as the eigenvalues appear as shifts in the characteristic speeds $c_{\text{s/f},q}$ and $a_q$ with respect to the velocity components $v_q$.

The numerical robustness and efficiency of the method was demonstrated by applying it to simulations of vortically driven nonlinear MHD waves traveling in two different gravitationally stratified plasmas in Section \ref{sec:level3}, inspired by \citetalias{Raboonik2024a}. Not only did the results show clear agreements with the well known core behaviors and properties of the slow, \alf, and fast MHD waves, it also shed light on how and why these properties exist the way they do. Moreover, mode conversion processes, which were previously understood only in the context of sub-3D linear MHD, were patently observable and readily measurable through the eigenenergies. We then investigated the numerical fidelity of the method and categorized possible sources of numerical inaccuracies. It was witnessed that even in the case of highly nonlinear disturbances in a structured 3D plasma model (Simulation II) which inevitably trigger various non-ideal-MHD numerical mechanisms (artificial diffusion), the method nonetheless performed with high precision (see Figure~\ref{fig:simIIEtot1D}).

In Section~\ref{sec:level2}, we discussed the main derivation of the EEDM equations based on the eigenmodes of the flux matrices $M_q$, in addition to a shortcut derivation approach by integrating the additional physics due to the forcing vector $\bs{S}$ into the pre-existing homogeneous base case derived in \citetalias{Raboonik2024b} using mathematical techniques. It is hoped that these instructions open up doors to incorporating additional physics beyond ideal-MHD into the EEDM. However, one difficulty is the strict mathematical requirement that any additional internal effects such as the Hall effect or ambipolar diffusion are contained in $M_q$, while any pure external forces must go into $\bs{S}$. This would throw the equations out of the quasi-linear (and consequently hyperbolic) form in the case of diffusive effects, which are modeled by second order partial spatial derivative terms. In other words, the mathematical ``tricks'' used in numerical solvers treating such internal effects as forcing terms would not work in the strictly mathematical framework of the EEDM, as any true internal effect would play into the characteristics/eigenmodes of the system, if any, while purely external effects do not.
Nevertheless, the current EEDM equations can still be applied to weakly non-ideal plasmas and highly conductive portions of stratified plasmas with confined non-ideal zones.

Finally, we invite researchers interested in either collaborative or independent work using the method to follow the open source GitHub repository  \href{https://github.com/raboonik/EEDM}{EEDM} for updates.

\section*{Acknowledgments}
This work was performed on the OzSTAR national facility at Swinburne University of Technology. The OzSTAR program receives funding in part from the Astronomy National Collaborative Research Infrastructure Strategy (NCRIS) allocation provided by the Australian Government, and from the Victorian Higher Education State Investment Fund (VHESIF) provided by the Victorian Government.
Computations were run on the Australian National Computational Infrastructure's Gadi machine through an award from Astronomy Australia Ltd.'s Astronomy Supercomputer Time Allocation Committee.
L.A.T. is supported by the National Solar Observatory.

\software{Analytic work performed in Wolfram Mathematica \citep{Mathematica}. Simulations performed in Lare 3.4.1 \citep{ARBER2001151}. Numerical analysis performed in AutoParallelizePy \citep{Raboonik_AutoParallelizePy_2024}, NumPy \citep{Harris:2020}, and SciPy \citep{Virtanen2020}, and figures were prepared using the Matplotlib library \citep{Hunter:2007}, all in Python3.}

%




\appendix
\section{Ideal-MHD Flux Matrices}\label{sec:AppendixA}
The flux matrices $M_q$ in each Cartesian direction $q\in\lrp{x,y,z}$ belonging to the ideal-MHD equations are given by
\begin{subequations}\label{eq:fluxMatrices}
    \begin{equation}
    M_x(\bs{x},t) = \begin{pmatrix}
        v_x & \rho & 0 & 0 & 0 & 0 & 0 & 0\\
        0   & v_x  & 0 & 0 & 0 & B_y/\mu_0\rho & B_z/\mu_0\rho & 1/\rho \\
        0 & 0 & v_x & 0 & 0 & -B_x/\mu_0\rho & 0 & 0 \\
        0 & 0 & 0 & v_x & 0 & 0 & -B_x/\mu_0\rho & 0 \\
        0 & 0 & 0 & 0 & v_x & 0 & 0 & 0 \\
        0 & B_y & -B_x & 0 & 0 & v_x & 0 & 0 \\
        0 & B_z & 0 & -B_x & 0 & 0 & v_x & 0 \\
        0 & \gamma p & 0 & 0 & 0 & 0 & 0 & v_x
    \end{pmatrix},
\end{equation}

\begin{equation}
    M_y(\bs{x},t) = \begin{pmatrix}
        v_y & 0 & \rho & 0 & 0 & 0 & 0 & 0\\
        0   & v_y  & 0 & 0 & -B_y/\mu_0\rho & 0 & 0 & 0 \\
        0 & 0 & v_y & 0 & B_x/\mu_0\rho & 0 & B_z/\mu_0\rho & 1/\rho \\
        0 & 0 & 0 & v_y & 0 & 0 & -B_y/\mu_0\rho & 0 \\
        0 & -B_y & B_x & 0 & v_y & 0 & 0 & 0 \\
        0 & 0 & 0 & 0 & 0 & v_y & 0 & 0 \\
        0 & 0 & B_z & -B_y & 0 & 0 & v_y & 0 \\
        0 & 0 & \gamma p & 0 & 0 & 0 & 0 & v_y
    \end{pmatrix},
\end{equation}

\begin{equation}
    M_z(\bs{x},t) = \begin{pmatrix}
        v_z & 0 & 0 & \rho  & 0 & 0 & 0 & 0 \\
 0 & v_z & 0 & 0 & -B_z/\mu_0\rho & 0 & 0 & 0 \\
 0 & 0 & v_z & 0 & 0 & -B_z/\mu_0\rho & 0 & 0 \\
 0 & 0 & 0 & v_z & B_x/\mu_0\rho & B_y/\mu_0\rho & 0 & 1/\rho \\
 0 & -B_z & 0 & B_x & v_z & 0 & 0 & 0 \\
 0 & 0 & -B_z & B_y & 0 & v_z & 0 & 0 \\
 0 & 0 & 0 & 0 & 0 & 0 & v_z & 0 \\
 0 & 0 & 0 & \gamma  p & 0 & 0 & 0 & v_z \\
    \end{pmatrix}.
\end{equation}
\end{subequations}

\section{Roe and Balsara Eigenmatrices}\label{sec:AppendixB}
The left eigenmatrices associated with each $M_q$, originally adapted from \cite{roe1996} and augmented to account for the divergence PA mode not considered therein, are given by
\begin{subequations}
    \begin{equation}
        L_x = \left(
\begin{array}{cccccccc}
 0 & 0 & 0 & 0 & 1 & 0 & 0 & 0 \\
 1 & 0 & 0 & 0 & 0 & 0 & 0 & -\frac{1}{c^2} \\
 0 & 0 & -\frac{\beta_{z \perp x}}{2} & \frac{\beta_{y \perp x}}{2} & 0 & -\frac{\beta_{z \perp x} \text{sgn}(a_x)}{2 \sqrt{\mu_0 \rho }} & \frac{\beta_{y \perp x} \text{sgn}(a_x)}{2 \sqrt{\mu_0 \rho }} & 0 \\
 0 & 0 & \frac{\beta_{z \perp x}}{2} & -\frac{\beta_{y \perp x}}{2} & 0 & -\frac{\beta_{z \perp x} \text{sgn}(a_x)}{2 \sqrt{\mu_0 \rho }} & \frac{\beta_{y \perp x} \text{sgn}(a_x)}{2 \sqrt{\mu_0 \rho }} & 0 \\
 0 & -\frac{\mathcal{C}_{\text{s},x}}{2 c^2} & -\frac{\mathcal{C}_{\text{f},x} \beta_{y \perp x} \text{sgn}(a_x)}{2 c^2} & -\frac{\mathcal{C}_{\text{f},x} \beta_{z \perp x} \text{sgn}(a_x)}{2 c^2} & 0 & -\frac{\alpha_{\text{f},x} \beta_{y \perp x}}{2 c \sqrt{\mu_0 \rho }} & -\frac{\alpha_{\text{f},x} \beta_{z \perp x}}{2 c \sqrt{\mu_0 \rho }} & \frac{\alpha_{\text{s},x}}{2 c^2 \rho } \\
 0 & \frac{\mathcal{C}_{\text{s},x}}{2 c^2} & \frac{\mathcal{C}_{\text{f},x} \beta_{y \perp x} \text{sgn}(a_x)}{2 c^2} & \frac{\mathcal{C}_{\text{f},x} \beta_{z \perp x} \text{sgn}(a_x)}{2 c^2} & 0 & -\frac{\alpha_{\text{f},x} \beta_{y \perp x}}{2 c \sqrt{\mu_0 \rho }} & -\frac{\alpha_{\text{f},x} \beta_{z \perp x}}{2 c \sqrt{\mu_0 \rho }} & \frac{\alpha_{\text{s},x}}{2 c^2 \rho } \\
 0 & -\frac{\mathcal{C}_{\text{f},x}}{2 c^2} & \frac{\mathcal{C}_{\text{s},x} \beta_{y \perp x} \text{sgn}(a_x)}{2 c^2} & \frac{\mathcal{C}_{\text{s},x} \beta_{z \perp x} \text{sgn}(a_x)}{2 c^2} & 0 & \frac{\alpha_{\text{s},x} \beta_{y \perp x}}{2 c \sqrt{\mu_0 \rho }} & \frac{\alpha_{\text{s},x} \beta_{z \perp x}}{2 c \sqrt{\mu_0 \rho }} & \frac{\alpha_{\text{f},x}}{2 c^2 \rho } \\
 0 & \frac{\mathcal{C}_{\text{f},x}}{2 c^2} & -\frac{\mathcal{C}_{\text{s},x} \beta_{y \perp x} \text{sgn}(a_x)}{2 c^2} & -\frac{\mathcal{C}_{\text{s},x} \beta_{z \perp x} \text{sgn}(a_x)}{2 c^2} & 0 & \frac{\alpha_{\text{s},x} \beta_{y \perp x}}{2 c \sqrt{\mu_0 \rho }} & \frac{\alpha_{\text{s},x} \beta_{z \perp x}}{2 c \sqrt{\mu_0 \rho }} & \frac{\alpha_{\text{f},x}}{2 c^2 \rho } \\
\end{array}
\right),
    \end{equation}

    \begin{equation}
        L_y = \left(
\begin{array}{cccccccc}
 0 & 0 & 0 & 0 & 0 & 1 & 0 & 0 \\
 1 & 0 & 0 & 0 & 0 & 0 & 0 & -\frac{1}{c^2} \\
 0 & -\frac{\beta_{z \perp y}}{2} & 0 & \frac{\beta_{x \perp y}}{2} & -\frac{\beta_{z \perp y} \text{sgn}(a_y)}{2 \sqrt{\mu_0 \rho }} & 0 & \frac{\beta_{x \perp y} \text{sgn}(a_y)}{2 \sqrt{\mu_0 \rho }} & 0 \\
 0 & \frac{\beta_{z \perp y}}{2} & 0 & -\frac{\beta_{x \perp y}}{2} & -\frac{\beta_{z \perp y} \text{sgn}(a_y)}{2 \sqrt{\mu_0 \rho }} & 0 & \frac{\beta_{x \perp y} \text{sgn}(a_y)}{2 \sqrt{\mu_0 \rho }} & 0 \\
 0 & -\frac{\mathcal{C}_{\text{f},y} \beta_{x \perp y} \text{sgn}(a_y)}{2 c^2} & -\frac{\mathcal{C}_{\text{s},y}}{2 c^2} & -\frac{\mathcal{C}_{\text{f},y} \beta_{z \perp y} \text{sgn}(a_y)}{2 c^2} & -\frac{\alpha_{\text{f},y} \beta_{x \perp y}}{2 c \sqrt{\mu_0 \rho }} & 0 & -\frac{\alpha_{\text{f},y} \beta_{z \perp y}}{2 c \sqrt{\mu_0 \rho }} & \frac{\alpha_{\text{s},y}}{2 c^2 \rho } \\
 0 & \frac{\mathcal{C}_{\text{f},y} \beta_{x \perp y} \text{sgn}(a_y)}{2 c^2} & \frac{\mathcal{C}_{\text{s},y}}{2 c^2} & \frac{\mathcal{C}_{\text{f},y} \beta_{z \perp y} \text{sgn}(a_y)}{2 c^2} & -\frac{\alpha_{\text{f},y} \beta_{x \perp y}}{2 c \sqrt{\mu_0 \rho }} & 0 & -\frac{\alpha_{\text{f},y} \beta_{z \perp y}}{2 c \sqrt{\mu_0 \rho }} & \frac{\alpha_{\text{s},y}}{2 c^2 \rho } \\
 0 & \frac{\mathcal{C}_{\text{s},y} \beta_{x \perp y} \text{sgn}(a_y)}{2 c^2} & -\frac{\mathcal{C}_{\text{f},y}}{2 c^2} & \frac{\mathcal{C}_{\text{s},y} \beta_{z \perp y} \text{sgn}(a_y)}{2 c^2} & \frac{\alpha_{\text{s},y} \beta_{x \perp y}}{2 c \sqrt{\mu_0 \rho }} & 0 & \frac{\alpha_{\text{s},y} \beta_{z \perp y}}{2 c \sqrt{\mu_0 \rho }} & \frac{\alpha_{\text{f},y}}{2 c^2 \rho } \\
 0 & -\frac{\mathcal{C}_{\text{s},y} \beta_{x \perp y} \text{sgn}(a_y)}{2 c^2} & \frac{\mathcal{C}_{\text{f},y}}{2 c^2} & -\frac{\mathcal{C}_{\text{s},y} \beta_{z \perp y} \text{sgn}(a_y)}{2 c^2} & \frac{\alpha_{\text{s},y} \beta_{x \perp y}}{2 c \sqrt{\mu_0 \rho }} & 0 & \frac{\alpha_{\text{s},y} \beta_{z \perp y}}{2 c \sqrt{\mu_0 \rho }} & \frac{\alpha_{\text{f},y}}{2 c^2 \rho } \\
\end{array}
\right),
    \end{equation}

    \begin{equation}
        L_z = \left(
\begin{array}{cccccccc}
 0 & 0 & 0 & 0 & 0 & 0 & 1 & 0 \\
 1 & 0 & 0 & 0 & 0 & 0 & 0 & -\frac{1}{c^2} \\
 0 & -\frac{\beta_{y \perp z}}{2} & \frac{\beta_{x \perp z}}{2} & 0 & -\frac{\beta_{y \perp z} \text{sgn}(a_z)}{2 \sqrt{\mu_0 \rho }} & \frac{\beta_{x \perp z} \text{sgn}(a_z)}{2 \sqrt{\mu_0 \rho }} & 0 & 0 \\
 0 & \frac{\beta_{y \perp z}}{2} & -\frac{\beta_{x \perp z}}{2} & 0 & -\frac{\beta_{y \perp z} \text{sgn}(a_z)}{2 \sqrt{\mu_0 \rho }} & \frac{\beta_{x \perp z} \text{sgn}(a_z)}{2 \sqrt{\mu_0 \rho }} & 0 & 0 \\
 0 & -\frac{\mathcal{C}_{\text{f},z} \beta_{x \perp z} \text{sgn}(a_z)}{2 c^2} & -\frac{\mathcal{C}_{\text{f},z} \beta_{y \perp z} \text{sgn}(a_z)}{2 c^2} & -\frac{\mathcal{C}_{\text{s},z}}{2 c^2} & -\frac{\alpha_{\text{f},z} \beta_{x \perp z}}{2 c \sqrt{\mu_0 \rho }} & -\frac{\alpha_{\text{f},z} \beta_{y \perp z}}{2 c \sqrt{\mu_0 \rho }} & 0 & \frac{\alpha_{\text{s},z}}{2 c^2 \rho } \\
 0 & \frac{\mathcal{C}_{\text{f},z} \beta_{x \perp z} \text{sgn}(a_z)}{2 c^2} & \frac{\mathcal{C}_{\text{f},z} \beta_{y \perp z} \text{sgn}(a_z)}{2 c^2} & \frac{\mathcal{C}_{\text{s},z}}{2 c^2} & -\frac{\alpha_{\text{f},z} \beta_{x \perp z}}{2 c \sqrt{\mu_0 \rho }} & -\frac{\alpha_{\text{f},z} \beta_{y \perp z}}{2 c \sqrt{\mu_0 \rho }} & 0 & \frac{\alpha_{\text{s},z}}{2 c^2 \rho } \\
 0 & \frac{\mathcal{C}_{\text{s},z} \beta_{x \perp z} \text{sgn}(a_z)}{2 c^2} & \frac{\mathcal{C}_{\text{s},z} \beta_{y \perp z} \text{sgn}(a_z)}{2 c^2} & -\frac{\mathcal{C}_{\text{f},z}}{2 c^2} & \frac{\alpha_{\text{s},z} \beta_{x \perp z}}{2 c \sqrt{\mu_0 \rho }} & \frac{\alpha_{\text{s},z} \beta_{y \perp z}}{2 c \sqrt{\mu_0 \rho }} & 0 & \frac{\alpha_{\text{f},z}}{2 c^2 \rho } \\
 0 & -\frac{\mathcal{C}_{\text{s},z} \beta_{x \perp z} \text{sgn}(a_z)}{2 c^2} & -\frac{\mathcal{C}_{\text{s},z} \beta_{y \perp z} \text{sgn}(a_z)}{2 c^2} & \frac{\mathcal{C}_{\text{f},z}}{2 c^2} & \frac{\alpha_{\text{s},z} \beta_{x \perp z}}{2 c \sqrt{\mu_0 \rho }} & \frac{\alpha_{\text{s},z} \beta_{y \perp z}}{2 c \sqrt{\mu_0 \rho }} & 0 & \frac{\alpha_{\text{f},z}}{2 c^2 \rho } \\
\end{array}
\right),
    \end{equation}
\end{subequations}
in which $\alpha_{\text{s/f},q}$ and $\beta_{q' \perp q}$ are dimensionless variables with the former defined immediately following Equations~\ref{eq:EEDMNew} and the latter given by Equation~\ref{eq:beta}. The variables $\mathcal{C}_{\text{s/f},q} = \alpha_{\text{s/f},q} c_{\text{s/f},q}$ are the slow/fast speeds scaled by their respective $\alpha$ variable.  

The corresponding right eigenvectors $R_q$ can be computed simply by inverting $L_q$ and simplifying the results using the following identities: $\alpha_{\text{s},q}^2 + \alpha_{\text{f},q}^2 = 1$, $\sum_{q' \neq q}\beta_{q'\perp q}^2 = 1$, $\alpha_{\text{s},q}^2 c_{\text{s},q}^2 + \alpha_{\text{f},q}^2 c_{\text{f},q}^2 = c^2$, and $c_{\text{s},q} c_{\text{f},q} = s_q c$.

\section{Drivers}\label{sec:AppendixC}
In this section, we briefly recapitulate the bottom wave drivers introduced in the Sections 3.2 and 4.2 of \citetalias{Raboonik2024a}. These are simply carried forward to this study and recycled in Section~\ref{sec:level3} using different parametric values specified here.

\subsection{Driver for Simulation I}\label{sec:AppendixC1}
Inspired by the linear torsional  solutions of \cite{Turkmani2004}, \citetalias{Raboonik2024a} (see the Equations 13 and 14 therein) fix the bottom boundary of their Simulations Ia and Ib to the following magnetic and velocity fields 
\begin{subequations}\label{eq:turkmani2004}
    \begin{equation}
        \bs{B}(z,t) = B_\perp(z,t)\lrp{\cos\lrp{k z - \omega t + \varphi},\sin\lrp{k z - \omega t + \varphi}, 0},
    \end{equation}
    \begin{equation}\label{eq:vperp}
        \bs{v} = -\frac{\bs{B}(z,t)}{B_0} a_0,
    \end{equation}
    \begin{equation}
        \omega = a_0 k,
    \end{equation}
    \begin{equation}
        B_\perp(z,t) = A \exp{\lrp{-\frac{(z - a_0 t - z_0)^2}{2 \sigma^2}}},
    \end{equation}
\end{subequations}
driving a vortical pulse.
In this study, $A = 10^{-3}$, $\sigma = 0.5$, $z_0 = -4.1556$,  $k = 2\pi / 5$, and $\varphi = \pi/4$ are the pulse's amplitude, width, initial vertical location, wavenumber, and phase, respectively. Note that $z_0$ is in the ghost cells (the main physical domain starts at $z = 0$; see Table~\ref{tab}) and its chosen value is to ensure a smooth introduction of the pulse into the simulation domain.

\subsection{Driver for Simulation II}\label{sec:AppendixC2}
To excite a torsional pulse around the magnetic spine of Simulation II, we will prescribe the following velocity and magnetic fields introduced by \citetalias{Raboonik2024a} (the Equations 16 and 17 therein; inspired by \cite{Galsgaard2003} and \cite{Turkmani2004})
\begin{subequations}\label{eq:galsDriver}
    \begin{equation}\label{eq:galsgaardVelocityBCs}
    \bs{v} = v_0 \exp\lrp{-\frac{x^2 + y^2}{2 \sigma_R^2}} \exp\lrp{-\frac{{\color{black} \lrp{t - t_0}^2}}{2 \sigma_{\text P}^2}} \lrp{y, -x, 0},
\end{equation}

\begin{equation}\label{eq:galsgaardFieldBCs}
    \bs{B} = \bs{B}_0 - \sqrt{\rho_0} \bs{v},
\end{equation}
\end{subequations}
wherein $\bs{B}_0$ is the initial magnetic field given by Equation~\ref{eq:galsgaardfield}, and 
$v_0 = 20$,
$\sigma_P = 0.07$,
$t_0 = 0.58179035$, and 
$\sigma_R = 0.1$ are respectively the velocity amplitude, vertical width, peak time, and horizontal width of the pulse.

\bibliographystyle{aasjournal}



\end{document}